\documentclass{article}[12pt]
\usepackage{authblk}
\usepackage{tabto} 
\usepackage{relsize} 
\usepackage{graphicx} 
\usepackage{tikz} 
\usepackage{float} 
\usetikzlibrary{calc,arrows}
\usepackage{amsmath,amssymb} 
\newcommand{\tarc}{\mbox{\large$\frown$}}
\newcommand{\arc}[2][-3ex]{{#2}{\kern #1{\raisebox{1.5ex}{\tarc}}}}
\usepackage{amsfonts} 
\usepackage[export]{adjustbox} 
\usepackage{algpseudocode}
\usepackage{algorithm}
\usetikzlibrary{calc,arrows}
\usepackage{todonotes}
\usepackage{comment}
\usepackage{wrapfig}
\usepackage{hyperref} 
\hypersetup{
    colorlinks,
    linkcolor={red!50!black},
    citecolor={green!50!black},
    urlcolor={blue!80!black}
}
\usepackage{siunitx} 
\usepackage{appendix}
\usepackage{multirow} 
\usepackage[
backend=biber,
style=alphabetic,
sorting=ynt
]{biblatex}
\usepackage{geometry}
\usepackage{array}
\usepackage{longtable}

\def\XXint#1#2#3{{\setbox0=\hbox{$#1{#2#3}{\int}$}
     \vcenter{\hbox{$#2#3$}}\kern-.5\wd0}}

\title{Comparison of inviscid and viscous vortex shedding from 
translating and rotating plates}

\author[1]{Yu Jun Loo \thanks{looyujun@umich.edu}}
\author[1]{Silas Alben}
\affil[1]{Department of Mathematics, University of Michigan, Ann Arbor, MI 48109, USA}
\date{}

\begin{document}

\maketitle
\begin{abstract}
    We compare an inviscid vortex sheet model with continuous leading-edge shedding with direct Navier-Stokes simulations over a wide range of unsteady plate motions at moderate Reynolds number ($\mathrm{Re} \approx 1000$). Approximately $70$ distinct kinematic configurations are examined, spanning both body-dominated and flow-dominated regimes. In body-dominated motions, where the fluid dynamics are primarily driven by prescribed plate accelerations, the inviscid model accurately reproduces normal force histories and the qualitative structure of the induced vorticity field. In flow-dominated configurations, with quasi-periodic vortex shedding, agreement with force predictions is good but reduced at low angles of attack, reflecting the greater sensitivity of vortex shedding dynamics to physical and computational parameters. The ability of the present formulation to accommodate stable, continuous leading-edge vortex shedding enables uniform comparisons across diverse motions and clarifies the regimes in which inviscid vortex sheet models can be used reliably for force prediction and physical interpretation.
\end{abstract}

\section{Introduction}
The remarkable efficiency of insect and bird flight has inspired a plethora of recent studies investigating propulsion by both rigid and flexible flapping plates \cite{pullin2004unsteady, nitsche2025stability, huang2016hovering, alben2021collective,alben2021collective2,hua2013locomotion,liu2020hydrodynamic}. These studies are motivated in part by their implications for the control and design of bio-inspired robots and micro air vehicles. In the same spirit, many authors have employed computational and experimental techniques to study the aerodynamic forces experienced by plates undergoing highly unsteady aerial maneuvers such as a complex pitch–ramp motion \cite{eldredge2009computational, yu2018experimental, ol2009high}. The relevant Reynolds numbers ($\mathrm{Re}$) of such maneuvers are typically $\sim 10^2$--$10^4$ \cite{eldredge2009computational}. Exploring large parameter spaces at these Reynolds numbers quickly becomes computationally prohibitive. As a result, many authors have turned to inviscid vortex sheet models as a computationally efficient alternative that can model and explain the aerodynamic mechanisms behind force generation during these maneuvers \cite{ramesh2014discrete,alben2012dynamics,huang2016hovering}. The physics of force generation on a body can often be explained by relating the unsteady forces to the vortex shedding patterns induced by the body \cite{alben2021collective}. Hence, by distilling the entire fluid–body system into a simplified vortex structure, vortex sheet models offer a promising route to both explaining and accurately predicting the aerodynamic forces experienced by plates and foils in terms of their vortex shedding patterns.
\newline

Several studies have compared inviscid vortex sheet models with direct Navier–Stokes simulations and/or experiments, in terms of the forces and vortex structures induced by solid bodies in flows \cite{nitsche1994numerical, darakananda2016vortex, xia2013lift, PanXiaoLE, SimulatingVortexFlapping, ramesh2014discrete, knowles2007integrated, FAURE201932}. In \cite{PanXiaoLE}, excellent agreement was obtained between the time-averaged thrust forces measured for a flapping airfoil in experiment and an inviscid vortex sheet model. A key caveat, however, is that good agreement was found only when leading-edge vortex shedding was included in the inviscid model. In both \cite{PanXiaoLE} and \cite{SimulatingVortexFlapping}, omitting leading-edge shedding from the inviscid model led to relatively poor agreement with the forces computed by a direct Navier–Stokes simulation.
\newline

These observations motivate a closer examination of leading-edge vortex shedding in vortex sheet simulations, which remains a central difficulty for such models. When leading-edge shedding occurs, free vorticity released from the body often remains very close to the body surface; by contrast, vorticity released at the trailing edge is quickly advected away. As a result, computing the velocity of vorticity released at the leading edge often involves a so-called “near-singular integral,” which can introduce severe numerical error \cite{nitsche2021evaluation} that persists in the inviscid setting due to the absence of viscous dissipation. Aside from these numerical issues, there is also a physical ambiguity. In inviscid flow, the boundary condition enforces no-penetration but not no-slip, so newly-shed vorticity must leave the body tangent to its edges. Hence, when the flow at a plate edge is directed onto the edge, it may not be clear to which side of the plate newly-shed vortex particles should be advected.
\newline

Many vortex sheet models therefore deliberately halt the shedding of vorticity at the leading edge when the flow is directed onto it \cite{albenFall, sohnFallingPlates, kansoHov, nitsche2025stability}. This is done especially frequently in coupled simulations, where the motion of the body is driven by the fluid forces and can become unstable in the presence of leading-edge shedding. When the motion of the body is prescribed (i.e., not coupled to the fluid forces), there is a related body of work surrounding the so-called “leading-edge suction parameter” (LESP): an experimentally determined quantity used in vortex sheet simulations to determine when leading-edge shedding should occur for bodies with rounded edges \cite{ramesh2014discrete,narsipur,ramesh2012theoretical}. This parameter depends on both the prescribed motion and the geometry of the body, and can be viewed as a modeling choice that allows one to fit vortex sheet simulations to experiments or direct Navier–Stokes simulations. In \cite{FAURE201932}, for example, the critical LESP value was chosen based on the body geometry and the desired $\mathrm{Re}$. In practice, both special quadrature rules \cite{nitsche2021evaluation, sohnFallingPlates} and smoothing techniques \cite{albenRegularize,alben2013efficient} are often necessary to resolve the near-leading-edge flow accurately and robustly in vortex sheet models.
\newline

Recently, a vortex sheet model has been proposed by the present authors that allows for stable, continuous leading-edge vortex shedding in vortex sheet simulations with strong flow-body coupling \cite{loo2025falling}. This model was used to simulate freely falling plates undergoing both side-to-side fluttering and end-over-end tumbling. Its predicted value of the plate density parameter where fluttering transitions to tumbling was consistent with both experiments and direct Navier–Stokes simulation \cite{loo2025falling}. Here, we build on that approach with an improved formulation and address a complementary but distinct question: to what extent can an inviscid vortex sheet model with continuous leading-edge shedding quantitatively reproduce forces and vortex dynamics from direct Navier–Stokes simulations?
\newline

Previous comparisons between vortex sheet models and viscous simulations have typically focused on a small number of representative kinematics, chosen to illustrate specific flow features or the capabilities of a given formulation. These studies have played an important role in establishing the physical relevance of inviscid formulations. At the same time, because treatments of leading-edge shedding were often sensitive to kinematics or relied on model-specific assumptions, extending such comparisons in a uniform way across a broad range of unsteady motions has remained challenging. The continuous leading-edge shedding framework introduced in \cite{loo2025falling} reduces these sensitivities and allows us to conduct a more systematic assessment. In the present study, we use a single vortex sheet model to compare force predictions with direct Navier–Stokes simulations for $\approx70$ distinct plate maneuvers, providing a broad evaluation of model performance across a wide range of conditions.
\newline

A meaningful comparison requires that the viscous simulations accurately represent the same zero-thickness geometry assumed by the vortex sheet model. Indeed, the flows induced by plates of zero and small but nonzero thickness can be quite different. In \cite{nuriev2019numerical}, for instance, the relative error between the time-averaged pressure drag on an oscillating plate with zero and nonzero thickness ($5\%$ of the plate length) at $\mathrm{Re} = 1000$ is as large as $\approx 15\%$. The pressure jump across a zero-thickness plate is greater at all times, consistent with the classical result that the normal force on the plate increases with decreasing thickness in an unbounded domain \cite{plotkin1981thickness}. A similar effect arises when the leading edge of a flat plate is rounded, which can change the normal force coefficient by about $25\%$, possibly by altering both the separation point and circulation intensity at the leading edge \cite{rival2014characteristic}. In many of the aforementioned studies comparing inviscid models with direct Navier–Stokes simulations, the zero-thickness plate is instead approximated by a plate of small but nonzero thickness ($\approx 2$–$6\%$ of the plate length).
\newline

At the same time, accurate Navier–Stokes simulation of a plate of exactly zero thickness presents well-known numerical challenges. The pressure and vorticity fields exhibit inverse-square-root singularities at the plate tips \cite{alben2021collective,Koumoutsakos_Shiels_1996}, and resolving these requires specialized numerical treatment. Prior studies have addressed these difficulties using approaches such as grid clustering near the tips \cite{alben2021collective, tamaddon1994unsteady, najjar1995simulations}, viscous particle methods that integrate vorticity to remove singular behavior \cite{Koumoutsakos_Shiels_1996}, and discrete kernel methods suitable for high-Reynolds-number immersed-plate problems \cite{sader2024starting}. Even with high-order methods and fine resolution, the computed forces can differ significantly depending on how the singular region is treated \cite{najjar1995simulations, xu2014scaling}. Following \cite{alben2021collective, tamaddon1994unsteady, najjar1995simulations}, we therefore cluster the grid near the plate tips to mitigate these singularities. 
\newline

In this study, we consider plates of exactly zero thickness and validate our viscous solver against the benchmark results of \cite{Koumoutsakos_Shiels_1996}. We then compare the aerodynamic forces and resulting vorticity fields for several unsteady plate maneuvers using both direct Navier–Stokes simulations and an improved vortex sheet model with continuous leading-edge shedding. This comparison will help elucidate which vortex-dynamical features of unsteady high-Re flows can be approximated well by inviscid models. Because inviscid vortex sheet simulations are both orders of magnitude faster and less expensive than corresponding viscous simulations, the present results also indicate where such models can be reliably used for rapid exploration of large kinematic parameter spaces or control.
\newline

The structure of the paper is as follows. In section \ref{methodology and validation section} we describe our viscous (section \ref{viscous numerical model subsection}) and inviscid (section \ref{inviscid numerical model subsection}) models. Section \ref{comparing inviscid and viscous simulations section} compares the forces and induced vorticity fields in five scenarios, namely plate 
oscillation (section \ref{oscillating plates subsection}), heaving and pitching (section \ref{heaving and pitching subsection}), pitch-up (section \ref{pitch up subsection}), uniform rotation (section \ref{uniform rotation subsection}) and steady translation (section \ref{steady translation subsection}). In section \ref{summary statistics section} we summarize our findings before we conclude in section \ref{conclusion section}.
\clearpage
\section{Methodology and validation}
\label{methodology and validation section}

Here we describe both the viscous and inviscid numerical models of the zero-thickness plate. In both models, we assume 2D flow, and the plate is represented by a line segment of unit length. The motion of the plate is completely characterized by two quantities, the position of its center of mass and its orientation angle (i.e.~the oriented angle a line tangent to the plate makes with the $x$-axis) denoted 
\begin{align*}
    \boldsymbol{X}_G(t) &: \text{plate center of mass} \\
  \beta(t) &: \text{plate orientation angle}
\end{align*}
 
\subsection{Viscous numerical model}
\label{viscous numerical model subsection}
We consider a plate with zero thickness and unit length occupying the interval $[-1/2,1/2]$ on the $x$-axis and solve the 2D Navier-Stokes equations in the body frame of the plate \cite{li2002moving} given by 

\begin{equation}
\begin{cases}
\partial_t \boldsymbol{u} + \boldsymbol{u}\cdot\nabla\boldsymbol{u} - \frac{1}{\mathrm{Re}}\Delta\boldsymbol{u} = -\nabla p - 2\boldsymbol{\Omega}\times\boldsymbol{u} - \boldsymbol{\Omega}\times(\boldsymbol{\Omega}\times \boldsymbol{r}) -\dot{\boldsymbol{\Omega}}\times\boldsymbol{r}  - \boldsymbol{A}(t)  \\
\nabla\cdot\boldsymbol{u} = 0
\end{cases}
\label{unsimplified Navier Stokes equation in the body frame of the plate}
\end{equation}
 A complete derivation of equation \eqref{unsimplified Navier Stokes equation in the body frame of the plate} is given in appendix \ref{navier stokes in body frame derivation appendix}. Here, $\boldsymbol{r} = [x,y,0]^T$, $\boldsymbol{u} = [u,v,0]^T$ is the fluid velocity, and $\boldsymbol{A}(t) = [A_x(t),A_y(t),0]^T$ is the acceleration of the plate center of mass (expressed in body-frame coordinates). $\boldsymbol{\Omega}(t) = [0,\ 0,\ \Omega(t)]^T$ is the vector of angular velocity of the plate about its center of mass, and $\dot{\boldsymbol{\Omega}}(t)=[0,\ 0,\ \dot \Omega(t)]^T$ is its angular acceleration vector. By definition, we have that $\frac{d}{dt}\beta(t) = \Omega(t)$. We nondimensionalize distances by the plate length $L$ and velocities by a characteristic velocity $U$ defined separately for each problem in section \ref{comparing inviscid and viscous simulations section}. Then $\mathrm{Re} = UL/\nu$ is a dimensionless inverse viscosity that we set to 1000. These equations can be simplified in terms of the variables $u,v,p$ yielding:
\begin{equation}
\begin{cases}
    \partial_t u + u\partial_xu + v\partial_yu - \frac{1}{\mathrm{Re}}\Delta u= -\partial_x p + 2\Omega v  + \dot{\Omega}y  +\Omega^2x - A_x,\\
    \partial_t v + u\partial_xv + v\partial_yv - \frac{1}{\mathrm{Re}}\Delta v= -\partial_y p -2\Omega u - \dot{\Omega}x + \Omega^2y - A_y, \\
    \partial_x u + \partial_y v = 0.
\end{cases}
\label{Navier Stokes equation in the body frame of the plate}
\end{equation}
Henceforth we let $\boldsymbol{r} = [x,y]^T$, $\boldsymbol{u} = [u,v]^T$ and $\boldsymbol{A}(t) = [A_x(t),A_y(t)]^T$. Since the centrifugal force $[\Omega(t)^2x,\Omega(t)^2 y]^T$ and the inertial force $[-A_x(t),-A_y(t)]^T$ are conservative, we may write them as the gradient of a potential:
\begin{equation}
    [\Omega(t)^2x - A_x(t),\Omega(t)^2y - A_y(t)]^T = \nabla(\frac{1}{2}\Omega(t)^2 (x^2 + y^2)- A_x(t)x - A_y(t)y).
\end{equation}
Hence, as the pressure is unique only up to the addition of a potential, we may absorb these forces into the pressure to form the simplified equations:
\begin{equation}
    \begin{cases}
    \partial_t u + u\partial_xu + v\partial_yu - \frac{1}{\mathrm{Re}}\Delta u= -\partial_x p + 2\Omega v  + \dot{\Omega}y,\\
    \partial_t v + u\partial_xv + v\partial_yv - \frac{1}{\mathrm{Re}}\Delta v= -\partial_y p -2\Omega u - \dot{\Omega}x, \\
    \partial_x u + \partial_y v = 0.
\end{cases}
\label{final set of equations equation}
\end{equation}
Note that only the Coriolis force $[2\Omega v,-2\Omega u]^T$ and the Euler force $[\dot\Omega y,-\dot\Omega x]^T$ remain, as these forces are non-conservative. 
\newline

Having presented the equations, we now briefly describe how we solve them numerically. Our viscous model for the flat plate is based on the method described in \cite{alben2021collective} with a few modifications. A similar model was also employed to study heat transfer from a flapping plate in \cite{Wang_Alben_2025} and validated at $ \mathrm{Re}\sim 150$. Here we briefly review some of the essential details. The system of equations \eqref{final set of equations equation} is solved on a rectangular grid with length $\ell_x$ and width $\ell_y $ in the $x$ and $y$ directions respectively. $x$ and $y$ are discretized with $m$ and $n$ points respectively. It is a finite-difference method, using a second-order stencil to compute spatial derivatives on a staggered MAC (marker-and-cell) grid aligned with the plate. The grid is non-uniform, and clustered near both the sides and edges of the plate. More precisely, assume we have (piecewise) uniformly spaced grids on $X$ and $Y$ such that $$-l_x/2 < X < l_x/2,\ -l_y/2 < Y < l_y/2.$$ Here $l_x$ and $l_y$ are the length and width respectively of the computational domain. The uniform grids are mapped to non-uniform grids on $(x, y)$ by:
\begin{equation}
    x = 
    \begin{cases}
        X  - \eta_L\frac{l_x - 1}{4\pi}\sin(4\pi (X - 1/2)/(l_x - 1)), \ \  \text{ if  }\  -l_x/2 < X < -1/2  \\
        X  + \ \eta\frac{1}{2\pi}\sin(2\pi  X),\  \  \ \ \ \ \ \ \ \ \ \ \ \ \ \ \ \ \ \ \ \ \ \ \ \ \text{  if  }\ \ \  -1/2 <X <\ 1/2 \\
        X - \eta_R \frac{l_x - 1}{4\pi}\sin(4\pi(X + 1/2)/(l_x- 1)),\ \ \ \text{if  }\ \ \ \ \ \ 1/2< X <  \ l_x/2 
    \end{cases}
\end{equation}
\begin{equation}
    y = Y + \eta\frac{l_y}{2\pi}\sin(\frac{2\pi}{l_y} Y
)\end{equation}
Here we take $l_y = l_x = 22$, and $\eta = 0.95$. $\eta_L$ and $\eta_R$ are chosen so that the mesh length changes smoothly across the edges of the plate (i.e.~the mesh length is a smooth function of $x$ at $\pm 1/2$). Concentrating the grid points near the edges of the plate is necessary to suppress the inverse-square-root singularities in both the vorticity and pressure fields and allow for accurate computation of aerodynamic forces. As discussed in \cite{alben2021collective}, these singularities can also pose an obstruction to convergence. The highly stretched grid does however raise the condition number of the linear system \cite{alben2021collective}. Since the Navier-Stokes equations are solved in the body frame of the plate, the $x$-axis of the grid is aligned with the plate regardless of its orientation angle. 
\newline

In all that follows, $X$ and $Y$ are discretized with $768$ points. The minimum and maximum mesh lengths are roughly $4\times10^{-4}$ and $6\times10^{-2}$ respectively. $dt = 2\times10^{-4}$ is chosen so that for each motion and corresponding characteristic velocity $U$ the maximum convective CFL number $\approx \frac{dt}{\min{dx}}U \leq 0.5 < 1$. The time derivative is discretized with second-order backward differentiation (BDF2) and we solve for the pressure iteratively using a variation of the SIMPLE scheme \cite{ferziger2019computational}.
To accelerate convergence we apply a 6th-order Shapiro filter \cite{falissard2013genuinely} before solving for the pressure at each iteration of the SIMPLE scheme. 

\subsubsection{Boundary conditions}

\begin{figure}[H]
    \centering
    \includegraphics[width=0.9\linewidth, trim = 1cm 4cm 1cm 4cm, clip]{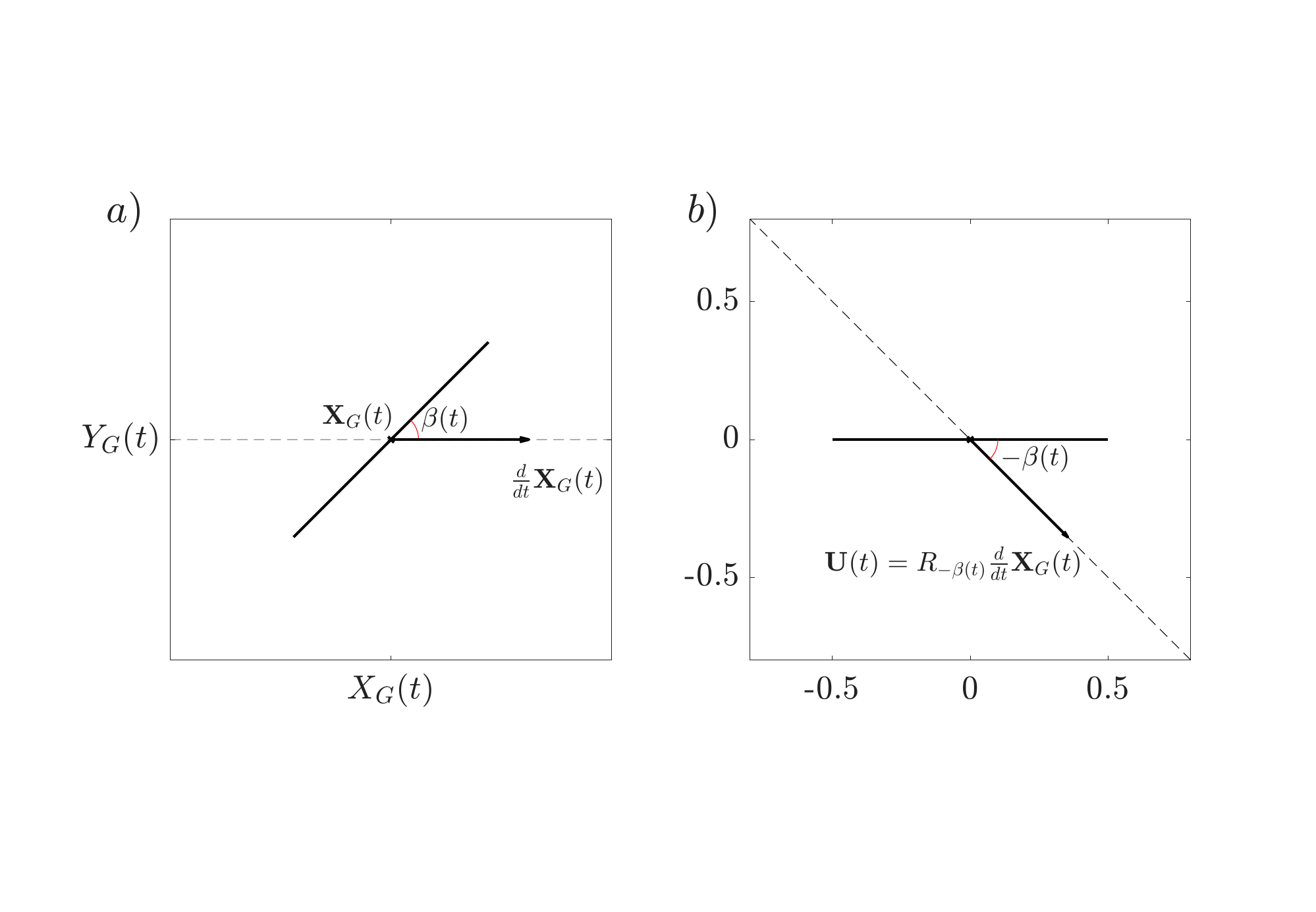}
    \caption{Schematic showing the plate in $(a)$ the lab frame and $(b)$ the body frame.}
    \label{fig:frame diagram figure}
\end{figure}

We seek to solve \eqref{final set of equations equation} on an unbounded domain. Assuming that the plate is initially at rest, the motion of the plate is completely determined by the translational velocity of the plate center of mass $\boldsymbol{U}(t)$ and its angular velocity $\Omega(t)$. To be precise, the translational velocity expressed in body-frame coordinates, $\boldsymbol{U}(t)$, is given by 
\begin{equation}
    \boldsymbol{U}(t) = R_{-\beta(t)} \frac{d}{dt}\boldsymbol{X}_G(t).
\end{equation}
The vector $\frac{d}{dt}\boldsymbol{X}_G(t)$ gives the translational velocity of the plate center in the laboratory frame. The rotation $R_{-\beta(t)}$ changes coordinates, expressing this velocity in the body frame.
 We illustrate this in figure \ref{fig:frame diagram figure}. In panel a we show an example of the plate in the lab frame, at position $\boldsymbol{X}_G(t) = [X_G(t),Y_G(t)]^T$ making orientation angle $\beta(t)$ with the $x$-axis and moving in the direction $\frac{d}{dt}\boldsymbol{X}_G(t)$. In panel b we show the corresponding situation in the body frame. The plate occupies $[-1/2,1/2]\times\{0\}$ and the velocity of the body expressed in the coordinates of the body frame must be rotated so that $\boldsymbol{U}(t) = R_{-\beta(t)}\frac{d}{dt}\boldsymbol{X}_G(t)$.

The reason for this definition is that for a plate moving with velocity $\frac{d}{dt}\boldsymbol{X}_G(t)$), in the body frame of the plate, the far field moves with velocity  
\begin{equation}
    \boldsymbol{U}_f(t,x,y) = [U_{f}(t,x,y),V_{f}(t,x,y)]^T =-\boldsymbol{U}(t) - [-\Omega(t) y, \Omega(t) x]^T.
\end{equation}
Let $\hat{\boldsymbol{S}}$ and $\hat{\boldsymbol{N}}$ denote the unit tangent vector and inward unit normal vector respectively along the computational boundary (with anticlockwise orientation). We use the far-field velocity $\boldsymbol{U}_f$ to impose Dirichlet and Neumann boundary conditions on the boundary where $\boldsymbol{U}_f$ is flowing into and out of the fluid domain respectively. 
The tangential velocity at the boundary is specified with Neumann boundary conditions. Stated more precisely, for each point in our computational boundary, we let 
\begin{align}
\begin{cases}
    \partial_{\hat{\boldsymbol{N}}}(\boldsymbol{u}-\boldsymbol{U}_f ) = 0\ &\text{ if $\boldsymbol{U}_f\cdot\hat{\boldsymbol{N}} < 0$} \\
    \partial_{\hat{\boldsymbol{N}}}(\boldsymbol{u} - \boldsymbol{U}_f)\cdot\hat{\boldsymbol{S}}  = 0,\ \text{\ } (\boldsymbol{u} - \boldsymbol{U}_f)\cdot \hat{\boldsymbol{N}}=0\ &\text{ if$\ \boldsymbol{U}_f\cdot\hat{\boldsymbol{N}} > 0$}.
\end{cases}    
\end{align}

To suppress any spurious vorticity that may be generated at the walls of the computational domain, we supplement our boundary conditions with an absorbing sponge layer using Rayleigh damping near the computational boundaries \cite{slinn1998model}. This approach has been used successfully to suppress boundary reflections in both incompressible \cite{pant2016viscous, slinn1998model} and compressible \cite{hu2008absorbing, hu2005perfectly} vortex-dominated flows in an unbounded domain. The application of this sponge boundary condition consists of adding a damping term 
\begin{equation}-\sigma(x,y)(\boldsymbol{u} -\boldsymbol{U}_f)
\end{equation}
to the right-hand side of \eqref{final set of equations equation} in a thin region (sponge layer) containing the boundary. Here we take the damping function to be 
\begin{equation}
    \sigma(x,y) = \begin{cases} 0.5((x^2 + y^2)^{ 1/2}-R)^{3}\ \text{  if } \ x^2 +y^2 > R^2\ \\ 0 \text{\ otherwise} \end{cases}
\end{equation} $\sigma(x,y)$ controls the strength and spatial distribution of the absorbing layer \cite{hu2008absorbing}, where $R=8$ defines the radius of our computational domain, so that our sponge layer has a minimum thickness of $\epsilon = 3$. Note that outside this circle, the true Navier-Stokes equations are not being solved, and hence we think of the region outside the circle as part of the boundary, not the computational domain.

\subsubsection{Validation}
\begin{figure}[H]
    \centering
    \includegraphics[width=0.9\linewidth]{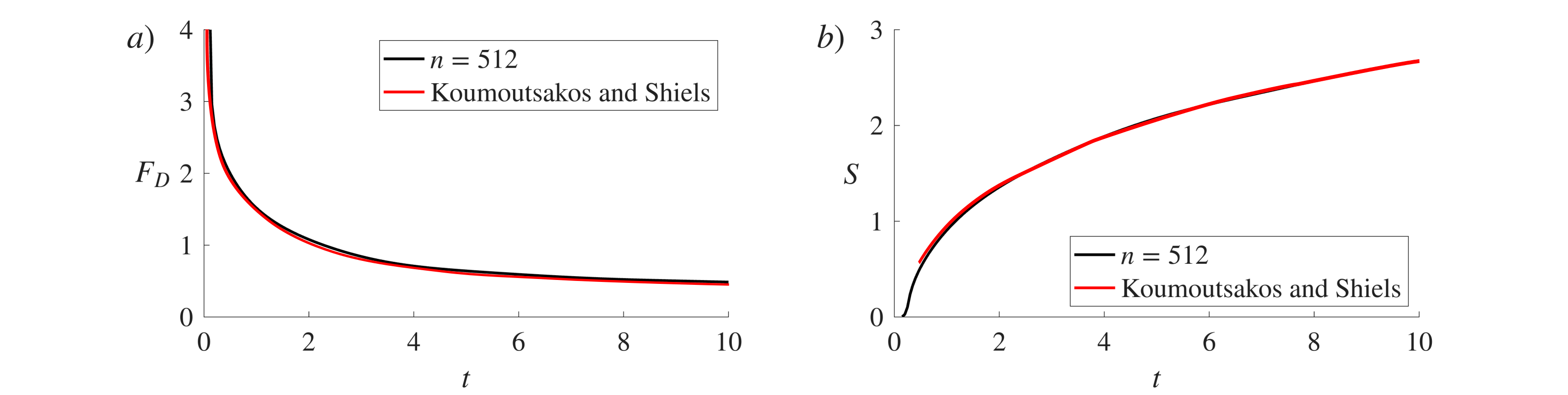}
    \caption{(a) The drag coefficient $F_D$ and (b) the length of the recirculation bubble $S$ over time for an impulsively started flat plate moving normal to itself with unit speed at $\mathrm{Re} = 1000$, compared with data obtained from figure 13 in \cite{Koumoutsakos_Shiels_1996}.}
    \label{drag coefficient comparison figure}
\end{figure}

To validate our viscous solver, we consider an impulsively started flat plate moving normal itself with unit speed at $\mathrm{Re} = 1000$. In figure \ref{drag coefficient comparison figure} we compare the computed forces between our viscous solver with the numerical results of \cite{Koumoutsakos_Shiels_1996}. The results are in good agreement. 

\begin{figure}[H]
    \centering
    \includegraphics[width=0.7\linewidth, trim = 0 0 0 0, clip]{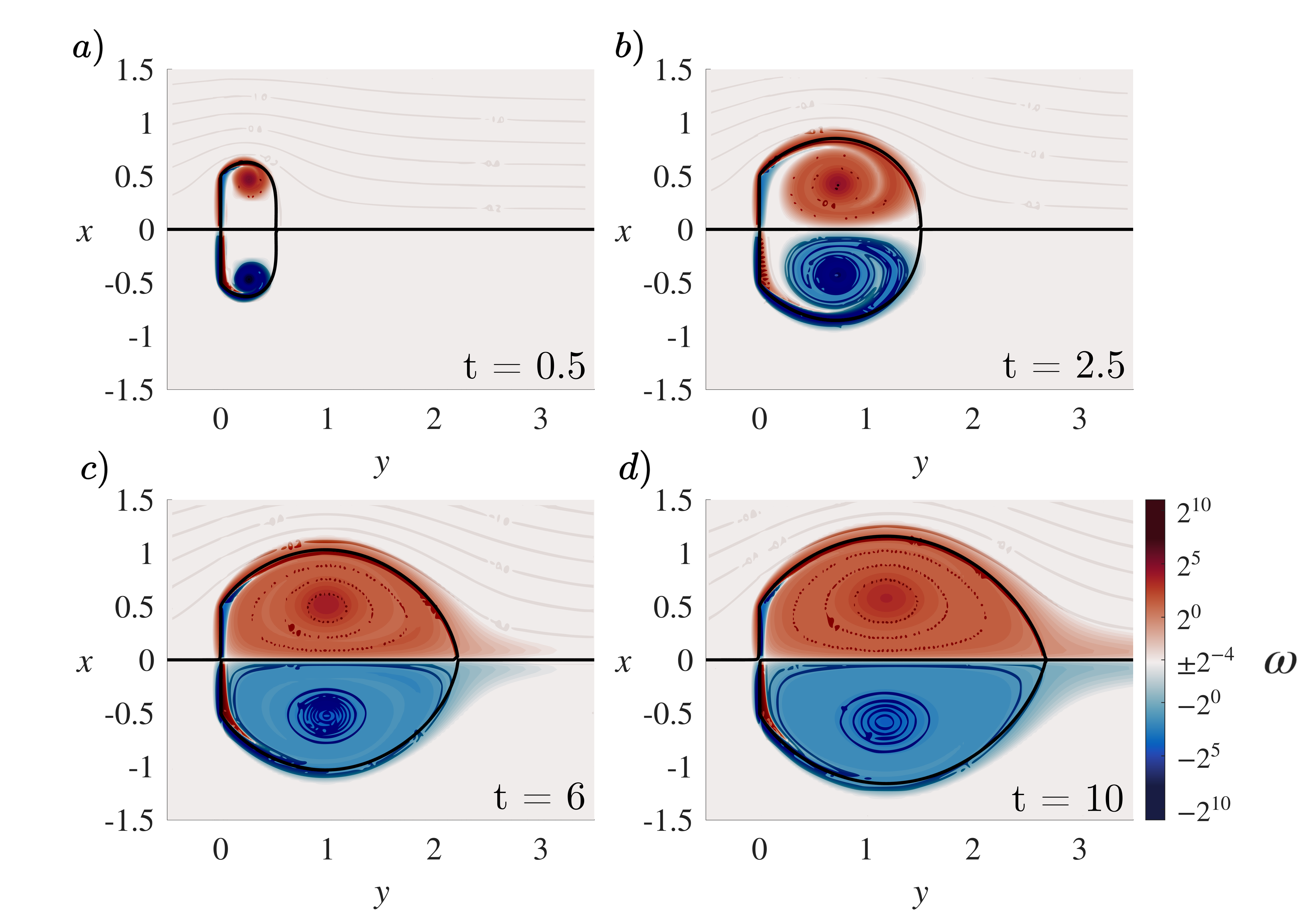}
    \caption{Vorticity fields (labeled by color bar) for an impulsively started plate at times $t$ = 0.5, 2.5, 6, and 10, with circulation bubble outlined in black and $\mathrm{Re} = 1000$. The streamlines (top half) and vorticity contours (bottom half) from figure 20 in \cite{Koumoutsakos_Shiels_1996} are overlaid on each panel.}
    \label{vorticity contours comparison figure}
\end{figure}

In figure \ref{vorticity contours comparison figure} we reproduce the vorticity contours shown in figure 20 of \cite{Koumoutsakos_Shiels_1996}. The circulation bubble is outlined in black, and our vorticity is presented on a bi-directional logarithmic scale, with contour levels ranging from $\pm2^{-4}$ to $\pm 2^{10}$. Here, the vorticity contours are visualized with the perceptually uniform color map ``balance" described in \cite{thyng2016true} (i.e.~each doubling of vorticity magnitude corresponds to a fixed decrease in color brightness). Each panel is overlaid with the corresponding streamlines (top half) and vorticity contours (bottom half) in figure 20 in \cite{Koumoutsakos_Shiels_1996}. In the top half of each panel the streamlines from \cite{Koumoutsakos_Shiels_1996} inside are given as red dotted lines, while in the bottom half of each panel, the vorticity contours are given as dark blue lines. The circulation bubble in the top half of the figure in \cite{Koumoutsakos_Shiels_1996} overlaps with the present viscous model.

\subsection{Inviscid numerical model}
\label{inviscid numerical model subsection}
We now describe the inviscid model. The model is essentially the same as that in \cite{loo2025falling}, but with a few improvements which we will highlight. The discussion presented here is similar to \cite{loo2025falling} but we use vector notation instead of complex variables. In the limit of zero viscosity, the two boundary layers along the two sides of the plate shrink to a zero-thickness vortex sheet along the body surface. In the limit of zero body thickness, these two vortex sheets merge into a single vortex sheet called the ``bound" vortex sheet, and the fluid velocity tangent to the body surface jumps discontinuously across the vortex sheet \cite{eldredge_inviscid_flows}.
\newline

In our zero-viscosity model there is no diffusion of vorticity from the vortex sheets, only advection from the separation points, which we assume to occur at the two edges of the plate only.  Consequently, two ``free" vortex sheets emanate from the body, one from each edge. The bound vortex sheet along the body and the free vortex sheets form one continuous vortex sheet that we compute at each time step.
\newline

As the body is accelerated into motion, there is a corresponding fluid motion that can be computed from the strengths and positions of the bound and free vortex sheets. As time increases, the lengths of the free vortex sheets increase at each edge, starting from zero. 
The bound vortex sheet coincides with the flat plate, given by $\boldsymbol{X}(s,t) = [X(s,t),Y(s,t)]^T, -1/2 \leq s \leq 1/2$. Note that 
\begin{equation}
\boldsymbol{X}(s,0) = [s,0]^T,\ -1/2 \leq s \leq 1/2 
\end{equation}
\begin{equation}
    \boldsymbol{X}(s,t) = \boldsymbol{X}_G(t) + {R}_{\beta(t)}\boldsymbol{X}(s,0)
\end{equation}  In this work, $\boldsymbol{X}_G(t)$ and $\beta(t)$ are prescribed, which determines $\boldsymbol{X}(s,t)$. The free vortex sheets emanating from the $s = \pm 1/2$ edges are denoted $\boldsymbol{X}_{+}(s,t) = [X_+(s,t),Y_+(s,t)]^T,\ s \in [1/2,s_{+} + 1/2]$ and $\boldsymbol{X}_{-}(s,t) = [X_-(s,t),Y_-(s,t)],\ s \in [-s_{-} -1/2,-1/2]$, respectively. Here, $s_{\pm}$ are the lengths of the positive and negative vortex sheets respectively, and the vortex sheet strengths $\gamma_{\pm}(s,t)$ are defined on the same intervals.
\newline

The strength of the entire (bound and free) vortex sheet is denoted $\gamma(s,t)$, with $s \in[-1/2 -s_-, 1/2 + s_+]$. We denote the restriction of $\gamma(s,t)$ to the negative, positive, and bound vortex sheets as $\gamma_-(s,t)$, $\gamma_+(s,t)$, and $\gamma_b(s,t)$ respectively.  
\newline

Convolving the vortex sheet strengths $\gamma_b$, and $\gamma_\pm$ with the Biot-Savart kernel $\boldsymbol{K}(\boldsymbol{r}) =\frac{1}{2\pi}\frac{[-y,x]^T}{|\boldsymbol{r}|^2}$ along the vortex sheet results in the following equation for the fluid velocity $\boldsymbol{u}$:
 \begin{multline}
    \boldsymbol{u}(\boldsymbol{r},t) = \int_{-1/2}^{1/2}{\boldsymbol{K}(\boldsymbol{r} - \boldsymbol{X}(s,t))\gamma_b(s,t)  ds}  \\+\int_{-s_{-} -1/2}^{-1/2}\boldsymbol{K}(\boldsymbol{r} - \boldsymbol{X_-}(s,t))\gamma_-(s,t) ds + \int_{1/2}^{s_{+} + 1/2}\boldsymbol{K}(\boldsymbol{r} - \boldsymbol{X}_+(s,t))\gamma_+(s,t) ds.
\end{multline}

Let $\Gamma(s,t)$ denote the arc length integral of $\gamma$ along the positive and negative vortex sheets, defined piecewise,
\begin{align}
\Gamma(s,t) = \begin{cases}
    \int_{-s_--1/2}^s \gamma(s',t)\,ds', \quad s \in [-s_--1/2, -1/2] \\
   \int_{1/2}^{s}\gamma_+(s',t)\,ds', \quad s \in [1/2, 1/2+s_+]
\end{cases} 
\end{align}

and $\Gamma_\pm(t)$ denote the total circulations in the positive and negative vortex sheets, i.e.
\begin{align}
\Gamma_+(t) \equiv \int_{1/2}^{1/2+s_{+}}\gamma_{+}(s,t)\, ds \quad ; \quad \Gamma_-(t) \equiv \int_{-s_{-}-1/2}^{-1/2}\gamma_{-}(s,t)\, ds.
\end{align}

Then, by reparameterizing the positions of the free vortex sheets in terms of $\Gamma$, we may rewrite the equation for the fluid velocity as

 \begin{multline}
    \boldsymbol{u}(\boldsymbol{r},t) = \int_{-1/2}^{1/2}{\boldsymbol{K}(\boldsymbol{r} - \boldsymbol{X}(s,t))\gamma_b(s,t)  ds}  \\+\int_{0}^{\Gamma_-(t)}\boldsymbol{K}(\boldsymbol{r} - \boldsymbol{X_-}(\Gamma,t))d\Gamma + \int_{0}^{\Gamma_+(t)}\boldsymbol{K}(\boldsymbol{r} - \boldsymbol{X}_+(\Gamma,t))d\Gamma.
    \label{unsmoothed fluid velocity equation}
\end{multline}

The first term on the right-hand side is the contribution of the bound vortex sheet, while the last two terms are the contributions of the negative and positive free vortex sheets respectively. Hence we see that the fluid velocity $\boldsymbol{u}$ can be determined from the unknown functions $\Gamma_\pm(t), \ \boldsymbol{X}_\pm(\Gamma,t)$, and $ \gamma_b(s,t)$. 
\newline

We now discuss how we solve for these functions. At each time step, $\Gamma_\pm(t)$ are chosen such that the Kutta conditions $\gamma_b(\pm1/2) = \gamma_\pm(\pm 1/2)$ are satisfied. In appendix \ref{kutta condition appendix} we discuss how this is done numerically. This condition ensures that the vortex sheet strength is finite and continuous across the edges of the plate. At each time step, $\boldsymbol{X_\pm}(t)$ is determined by solving the Birkhoff-Rott equations, which asks that the free sheets move with the average of the fluid velocity on the two sides of the vortex sheet. Hence from \eqref{unsmoothed fluid velocity equation} we obtain the classical Birkhoff-Rott equations:
\begin{multline}
    \partial_t\boldsymbol{X}_\pm(\Gamma,t) = \int_{-1/2}^{1/2}{\boldsymbol{K}\big(\boldsymbol{X}_\pm(\Gamma,t) - \boldsymbol{X}(s,t)\big)\gamma_b(s,t)  ds}  \\+\int_{0}^{\Gamma_-(t)}\boldsymbol{K}\big(\boldsymbol{X}_\pm(\Gamma,t) - \boldsymbol{X_-}(\Gamma',t)\big)d\Gamma' + \int_{0}^{\Gamma_+(t)}\boldsymbol{K}\big(\boldsymbol{X}_\pm(\Gamma,t) - \boldsymbol{X}_+(\Gamma',t)\big)d\Gamma'.
    \label{classical birkhoff rott equations equation}
\end{multline}

The classical Birkhoff-Rott equations (with principal valued integrals for the free sheets) are in fact ill-posed, and develop a curvature singularity at a finite critical time \cite{moore1979spontaneous, Krasny_1986, Shelley_1992}. Krasny showed that the Birkhoff-Rott equations can be evolved numerically past this critical time by introducing a smoothing parameter $\delta$ to ``desingularize" the equations \cite{krasny1986desingularization}. This regularization consists of replacing the kernel $\boldsymbol{K}(\boldsymbol{r})$ with $\boldsymbol{K}_\delta(\boldsymbol{r}) = \boldsymbol{K}(\boldsymbol{r})\frac{|\boldsymbol{r}|^2}{\boldsymbol{|r|^2 + \delta^2 }}$ in the free-vortex-sheet terms, eliminating the integrand singularity. This smoothing parameter inhibits the growth of small spatial features and prevents the formation of curvature singularities on free vortex sheets. However, the smoothing parameter jumps discontinuously from $\delta$ on the free vortex sheets to $0$ on the bound vortex sheet, which creates logarithmic singularities at the edges of the plate in the resulting flow velocity. Following the method described in \cite{loo2025falling}, it is possible to remove these logarithmic singularities by regularizing the contribution of the bound vortex sheet, i.e. by replacing the kernel $\boldsymbol{K}(\boldsymbol{r} - \boldsymbol{X})$ with $$\tilde{\boldsymbol{K}}_\delta(\boldsymbol{r},\boldsymbol{X}(\cdot),s) = B_\delta(\min_{s'}|\boldsymbol{r} - \boldsymbol{X}(s')|)\boldsymbol{K}(\boldsymbol{r} - \boldsymbol{X}(s)) + (1 - B_\delta(\min_{s'}|\boldsymbol{r} - \boldsymbol{X}(s')|))\boldsymbol{K}_\delta(\boldsymbol{r} - \boldsymbol{X}(s))$$ where $$B_\delta(l) = \begin{cases}
          0 \text{,\  if }l = 0, \\
          \frac{\exp(-\delta/l)}{\exp(-\delta/l) + \exp(-\delta/(\delta -l))} \text{,\  if }0 < l < \delta, \\
          1  \text{,\  if $l \geq \delta$}.
         \end{cases}$$ 

Note that $\min_{s'}|\boldsymbol{r} - \boldsymbol{X}(s')|$ is the distance from $\boldsymbol{r}$ to the curve parameterized by $\boldsymbol{X}$. Intuitively, $\tilde{\boldsymbol{K}}_\delta$ interpolates between the unsmoothed and smoothed kernels ($\boldsymbol{K}$ and $\boldsymbol{K}_\delta$ respectively) over a $\delta-$neighborhood of the plate. That is, $\tilde{\boldsymbol{K}}\approx \boldsymbol{K}_\delta$ when $\min_{s'}|\boldsymbol{r} - \boldsymbol{X}(s')| < \delta$, and $\tilde{\boldsymbol{K}}= \boldsymbol{K}$ when $\min_{s'}|\boldsymbol{r} - \boldsymbol{X}(s)| \geq \delta$. To summarize, we determine the position of the free vortex sheets $\boldsymbol{X}_\pm$ by solving the (blob-regularized) velocity-smoothed Birkhoff-Rott equations given by 
\begin{multline}
    \partial_t\boldsymbol{X}_\pm(\Gamma,t) = \int_{-1/2}^{1/2}{\tilde{\boldsymbol{K}}_\delta\big(\boldsymbol{X}_\pm(\Gamma,t), \boldsymbol{X}(\cdot,t),s\big)\gamma_b(s,t)  ds}  \\+\int_{0}^{\Gamma_-(t)}\boldsymbol{K}_\delta\big(\boldsymbol{X}_\pm(\Gamma,t) - \boldsymbol{X_-}(\Gamma',t)\big)d\Gamma' + \int_{0}^{\Gamma_+(t)}\boldsymbol{K}_\delta\big(\boldsymbol{X}_\pm(\Gamma,t) - \boldsymbol{X}_+(\Gamma',t)\big)d\Gamma'.
    \label{velocity smoothed birkhoff rott equation}
\end{multline}

$\gamma_b$ is determined by solving the no-penetration condition, which requires that the fluid velocity normal to the plate equal the plate's normal velocity, at the plate:
\begin{multline}
\hat{\boldsymbol{n}}\cdot\partial_t\boldsymbol{X} = \hat{\boldsymbol{n}}\cdot\bigg(\int_{-1/2}^{1/2}{\boldsymbol{K}\big(\boldsymbol{X}(s,t) - \boldsymbol{X}(s',t)\big)\gamma_b(s',t)  ds'}  \\+\int_{0}^{\Gamma_-(t)}\boldsymbol{K}_{\tilde{\delta}(s')}\big(\boldsymbol{X}(s,t) - \boldsymbol{X_-}(\Gamma',t)\big)d\Gamma' + \int_{0}^{\Gamma_+(t)}\boldsymbol{K}_{\tilde{\delta}(s')}\big(\boldsymbol{X}(s,t) - \boldsymbol{X}_+(\Gamma',t)\big)d\Gamma'\bigg).
    \label{no penetration equation}
\end{multline}

There are two noteworthy aspects of equation \eqref{no penetration equation} that we highlight. The first is that, unlike in (\ref{velocity smoothed birkhoff rott equation}), there is no smoothing of the body's contribution to the fluid velocity, which is given by the singular integral operator $C[\gamma_b](\boldsymbol{r}) = \int_{-1/2}^{1/2}\hat{\boldsymbol{n}}\cdot \boldsymbol{K}(\boldsymbol{r} - \boldsymbol{X}(s,t))\gamma_b(s,t)ds$. This because replacing the singular kernel in $C[\cdot]$ with a smooth function makes solving for $\gamma_b$ ill-posed. Indeed, consider the smoothed operator $C_\delta[\gamma_b] = \int_{-1/2}^{1/2}\hat{\boldsymbol{n}}\cdot \boldsymbol{K}_\delta(\boldsymbol{r} - \boldsymbol{X}(s,t))\gamma_b(s,t)ds$. Unlike $\boldsymbol{K}$, $\boldsymbol{K}_\delta$ is a smooth function. As an integral operator with smooth kernel, it is easy to show that $C_\delta$ is a compact operator from $L^2([-1/2,1/2]) \rightarrow L^2([-1/2,1/2])$. A compact operator cannot have a bounded inverse. Indeed, if it did have a compact inverse, the identity map, being the composition of a bounded operator and a compact operator, must be compact as well. But this contradicts the well-known fact that the unit ball is not compact in an infinite dimensional space. Hence, as the inverse of $C_\delta$ is unbounded, solving for $\gamma_b$ in the blob-regularized no-penetration condition is ill-posed for any $\delta > 0$, however small; as the inverse of $C_\delta$ is discontinuous, any small numerical error in the no-penetration condition would be magnified in $\gamma_b$ and prohibit any guarantee of accuracy or stability.
\newline

By contrast, \eqref{no penetration equation} can be solved by inverting the singular integral operator $C[\gamma_b](\boldsymbol{r}) =\int_{-1/2}^{1/2}\hat{\boldsymbol{n}}\cdot \boldsymbol{K}(\boldsymbol{r} - \boldsymbol{X}(s,t))\gamma_b(s,t)ds$ in the no-penetration condition. This determines $\gamma_b(s,t)$ to lie within a one-dimensional space of functions, each corresponding to a different amount of circulation around the plate. The particular value of circulation is set by Kelvin's circulation theorem, which requires that the total circulation in the flow is conserved. Since the plate is accelerated from rest, the initial circulation in the flow is zero. Hence Kelvin's circulation theorem dictates that 
\begin{equation}
    \int_{-1/2}^{1/2}\gamma_b(s,t)ds  +\Gamma_+(t) + \Gamma_-(t) = 0.
\end{equation} 
With this additional constraint, we may invert the operator $C[\cdot]$ to obtain a unique $\gamma_b(s,t)$. This is discussed in detail in appendix \ref{kutta condition appendix}.
\newline

The second important aspect of \eqref{no penetration equation} that we now highlight is that the smoothing parameter on the free sheets varies with $s' = s'(\Gamma')$, the arc length parameter on the free sheet. I.e. $\delta = \tilde{\delta}(s')$. More specifically, on the free vortex sheet connected the $\pm1/2$ edge of the plate, we take 
\begin{equation}
    \tilde{\delta}(s) =\delta_0  + (\delta - \delta_0) (1 - e^{-(( \pm 1/2 -s)/\delta)^2}).
\end{equation} 
Here we take $\delta_0 =\delta/1000$, a much smaller amount of regularization than $\delta$. Note that $s' = s'(\Gamma')$ is a function of the circulation. This aspect of the model is different from that of \cite{loo2025falling}, where a uniform regularization parameter was used. With $\tilde{\delta}$ tapered to a smaller value near the edge of the plate, the computed $\gamma_b$ more closely approximates the unregularized version \cite{albenRegularize, loo2025falling}. We also find that it reduces the sensitivity of the model to the far-field value of the smoothing parameter $\delta$, since near the edges of the plate $\tilde{\delta}(s) \approx 0$ regardless of the value of $\delta$. 
\newline

In summary, in order to remove the logarithmic singularities on the edges of the plate, we use both mismatched kernels ($\tilde{\boldsymbol{K}}_\delta$ and $\boldsymbol{K}$), and mismatched smoothing parameters ($\delta$ and $\tilde{\delta}(s')$)  in the Birkhoff-Rott equation \eqref{velocity smoothed birkhoff rott equation} and the no-penetration condition \eqref{no penetration equation}. As a result, when the free sheets are evolved, they are advected with a fluid velocity different from that which satisfies the no-penetration condition. This mismatch introduces an $O(\delta)$ error into the no-penetration condition that can cause the free sheets to penetrate the plate when solving for $\boldsymbol{X}_\pm(s,t)$. As shown in \cite{loo2025falling}, this error can be controlled by fencing: clipping the displacements of points normal to the plate to prevent them from passing through the plate at each sub-step of a time step. Fencing suppresses the error incurred from the mismatched velocities, resulting in a numerical scheme that was shown to be at least first order in time for the dynamics of falling plates with different densities computed in \cite{loo2025falling}. In general, the fluid dynamics are less complicated in the present work because the plate motion is fully prescribed. 
\newline

In table \ref{tableEqns} we list the coupled system of equations we solve in the first column, with the corresponding unknown variables in the second column.

\begin{longtable}{|>{\centering\arraybackslash}m{10cm}|>{\centering\arraybackslash}m{3cm}|}
\caption{Table showing the equations and unknown variables we solve for at each time $t$.} \label{tableEqns}\\
\hline
 \textbf{Equation} & \textbf{Unknown variable} \\
\hline
\begin{multline}
    \hat{\boldsymbol{n}}\cdot\partial_t\boldsymbol{X} = \hat{\boldsymbol{n}}\cdot\bigg(\int_{-1/2}^{1/2}{\boldsymbol{K}\big(\boldsymbol{X}(s,t) - \boldsymbol{X}(s',t)\big)\gamma_b(s',t)  ds'}  \\+\int_{0}^{\Gamma_-(t)}\boldsymbol{K}_{\tilde{\delta}(s')}\big(\boldsymbol{X}(s,t) - \boldsymbol{X_-}(\Gamma',t)\big)d\Gamma' + \\ \int_{0}^{\Gamma_+(t)}\boldsymbol{K}_{\tilde{\delta}(s')}\big(\boldsymbol{X}(s,t) - \boldsymbol{X}_+(\Gamma',t)\big)d\Gamma'\bigg)
    \label{no penetration table equation}
\end{multline}

\begin{equation} \label{kelvin table}
    \Gamma_+(t) + \Gamma_-(t) + \int_{-1/2}^{1/2}\gamma_b(s,t)ds = 0
\end{equation} 
&
$\gamma_b(s,t)$
\\
\hline
\begin{equation} \label{kutta condition table equation}
    \gamma_b(\pm1/2) = \gamma_\pm(\pm 1/2)
\end{equation}
&
$\Gamma_\pm(t)$
\\
\hline
\begin{multline}
    \partial_t\boldsymbol{X}_\pm(\Gamma,t) = \int_{-1/2}^{1/2}{\tilde{\boldsymbol{K}}_\delta\big(\boldsymbol{X}_\pm(\Gamma,t), \boldsymbol{X}(\cdot,t),s\big)\gamma_b(s,t)  ds}  \\+\int_{0}^{\Gamma_-(t)}\boldsymbol{K}_\delta\big(\boldsymbol{X}_\pm(\Gamma,t) - \boldsymbol{X_-}(\Gamma',t)\big)d\Gamma' \\ + \int_{0}^{\Gamma_+(t)}\boldsymbol{K}_\delta\big(\boldsymbol{X}_\pm(\Gamma,t) - \boldsymbol{X}_+(\Gamma',t)\big)d\Gamma'
    \label{velocity smoothed birkhoff rott table equation}
\end{multline}
& 
$\boldsymbol{X}_\pm(s,t)$\\
\hline
\end{longtable}

The fluid velocity can then be obtained from the following formula: 
\begin{multline}
    \boldsymbol{u}(\boldsymbol{r},t) = \int_{-1/2}^{1/2}{\tilde{\boldsymbol{K}}_\delta(\boldsymbol{r}, \boldsymbol{X}(\cdot,t),s)\gamma_b(s,t)  ds}  \\+\int_{0}^{\Gamma_-(t)}\boldsymbol{K}_\delta(\boldsymbol{r} - \boldsymbol{X_-}(\Gamma,t))d\Gamma + \int_{0}^{\Gamma_+(t)}\boldsymbol{K}_\delta(\boldsymbol{r} - \boldsymbol{X}_+(\Gamma,t))d\Gamma.
    \label{velocity equation}
\end{multline}

\subsubsection{Inviscid algorithm}
In the algorithm block below we summarize the numerical procedure that we use to solve for $\boldsymbol{X}_\pm(s,t),\ \Gamma_\pm(t),\ \gamma_b(s,t)$ simultaneously at each time step. 

\begin{algorithm}[H]
\caption{Inviscid vortex shedding algorithm when the motion of the body $\boldsymbol{X}(s,t)$ is known} \label{coupled fluid body solver}
\begin{algorithmic}[0]
\For{each time step $k = 1, 2, \ldots$}
    \State \textbf{Step 1:} Solve for the free-sheet positions $\boldsymbol{X}_\pm(s,t)$ explicitly
    
    \State \quad $\bullet$ Use the velocity-smoothed Birkhoff-Rott equations \eqref{velocity smoothed birkhoff rott table equation} to compute $\boldsymbol{X}_\pm(s,t)$ at the current time \\
    \quad\quad\quad\ step $t = dt\cdot k$ using 3rd-order Adams-Bashforth.
    \\
    \State \textbf{Step 2:} Sub-step fence for the free sheets.
    \State \quad $\bullet$ Project all crossing free-sheet points normally to the plate surface to eliminate crossing.
    \\
    \State \textbf{Step 3:} Create new points on the free sheets
    \State \quad $\bullet$ Update $\boldsymbol{X}(s,t)$ with the known value, and again project any crossing free-sheet points normal to \\
\quad\quad\quad\ the plate to eliminate crossing. \\
    \quad\quad\quad\ Add a new point to each of the free sheets, with positions exactly at the edges of the plate. \\
    \quad\quad\quad\ These new points are labeled with the new total circulation $\Gamma_\pm(t)$ of their respective sheets,
    \\
    \quad\quad\quad\ which we solve for in the next step.
    \\
    \State \textbf{Step 4:} Solve for the new free-sheet circulations $\Gamma_\pm(t)$ implicitly. 
    \State \quad $\bullet$  Construct the helper function $F(\Gamma_+,\Gamma_-) = [\gamma_b(1/2) - \gamma_+(1/2),\ \gamma_b(-1/2) - \gamma_-(-1/2)]^T$. \\
    \quad\quad\quad\ $F$ obtains $\gamma_b(\pm1/2)$ by inverting the singular integral in the no-penetration condition \eqref{no penetration table equation} \\
    \quad\quad\quad\ with Kelvin's circulation theorem \eqref{kelvin table} to fix the gauge. In appendix \ref{kutta condition appendix} we discuss how this is done. \\
    \quad\quad\quad\ Solve $F(\Gamma_+,\Gamma_-) = \boldsymbol{0}$ with Broyden's method to obtain $\Gamma_\pm$ and $\gamma_b(s,t)$ simultaneously. The initial \\ 
    \quad\quad\quad\ guess is obtained with second-order extrapolation in time.
\EndFor
\end{algorithmic}
\label{full numerical algorithm}
\end{algorithm}

\subsubsection{Computing aerodynamic forces}
By considering the Bernoulli equation in the reference frame of the moving body, and taking its difference when evaluated on both sides of the body, the following pressure-jump equation can be obtained \cite[85]{eldredge_inviscid_flows} \cite{point_vortex_attraction}:
\begin{equation}
    [p]^+_-(s,t) = \partial_t\big(\Gamma_-(t) + \Gamma_b(s,t)\big) + \gamma_b(s,t)(\boldsymbol{u}(\boldsymbol{X}(s,t),t)-\partial_t\boldsymbol{X}(s,t))\cdot\hat{\boldsymbol{s}}. \label{Bernoulli}
\end{equation}

Here, $[p]^+_-(s,t)$ is the pressure jump across the plate and $\Gamma_b(s,t) = \int_{-1/2}^s\gamma_b(s,t)\, ds$. The net normal force on the (inviscid) plate is therefore 
\begin{equation}
    C_N(t) = \int_{-1/2}^{1/2}[p]_{-}^+(s,t)ds.
\end{equation}

\subsubsection{Computing the vorticity}
The Biot-Savart law states that the velocity induced by a blob-regularized point vortex with circulation $\Gamma_0$ situated at position $\boldsymbol{x}_0$ is given by 
\begin{equation}
    \frac{1}{2\pi}\frac{(\boldsymbol{x} - \boldsymbol{x}_0)^\perp\Gamma_0}{|\boldsymbol{x}- \boldsymbol{x}_0|^2 + \delta^2}.
\end{equation} 
Following \cite{holm2006euler}, the blob-regularized vorticity induced by this point vortex is
\begin{equation}
    \omega_\delta(\boldsymbol{x}) = \boldsymbol{\nabla}\times\frac{1}{2\pi}\frac{(\boldsymbol{x} - \boldsymbol{x}_0)^\perp\Gamma_0}{|\boldsymbol{x}- \boldsymbol{x}_0|^2 + \delta^2} = \frac{\delta^2\Gamma_0}{\pi(|\boldsymbol{x}- \boldsymbol{x}_0|^2 + \delta^2)^2}
\end{equation}
For the case of a vortex sheet with length $L$ and cumulative circulation $\Gamma(s),\ 0 < s < L$ and position parameterized by $\boldsymbol{x}'(\Gamma(s))$, we may let $W_\delta(\boldsymbol{x}) = \frac{\delta^2}{\pi(|\boldsymbol{x}|^2 + \delta^2)^2}$ so that 
\begin{equation}
    \omega_\delta(\boldsymbol{x})= \int_0^{\Gamma(L)}W_\delta(\boldsymbol{x} - \boldsymbol{x}'(\Gamma'))d\Gamma'.
\end{equation}

\section{Comparing inviscid and viscous simulations}
In this section we compare the inviscid and viscous models described in the previous section across five distinct motions, some taken from previous studies: periodic oscillation (section \ref{oscillating plates subsection}), heaving and pitching in a background flow (section \ref{heaving and pitching subsection}), pitch-up (section \ref{pitch up subsection}), uniform rotation (section \ref{uniform rotation subsection}), and steady translation (section \ref{steady translation subsection}). The cases of steady translation and uniform rotation with no background flow have flow-dominated vortex dynamics while the others have body-dominated vortex dynamics. We classify the vortex dynamics as body-dominated when the free-vortex-sheet motions and resulting aerodynamic forces correspond to the body accelerations, so that the instantaneous forces are relatively insensitive to the evolution of the previously shed vorticity. Conversely, the vortex dynamics is referred to as flow-dominated when the body undergoes little or no acceleration and the forces are governed primarily by the natural dynamics of the free vortex sheets, such as quasi-periodic vortex shedding. 
\newline

All results in this section are shown for $\mathrm{Re}=1000$. To assess whether the observed level of agreement is specific to this Reynolds number, appendix \ref{reynolds number sensitivity appendix} presents a representative comparison for $\mathrm{Re}=500$, $1000$, and $2000$.

\label{comparing inviscid and viscous simulations section}

\subsection{Oscillating plates}
\label{oscillating plates subsection}
First, we consider a plate oscillating normal to itself with frequencies $\omega = \frac{2\pi}{KC}$ where $\textrm{KC} = 0.4, 0.8, \ldots, 3.6$. That is, 
\begin{align}\boldsymbol{X}_G(t) &= \frac{\textrm{KC}}{2\pi}\sin(\frac{2\pi}{\textrm{KC}}t)[0,1]^T, \\ \beta(t) &= 0. 
\end{align} 
Here, KC is the dimensionless period. The amplitude is proportional to KC, ensuring that $\max_t |d\boldsymbol{X}_G(t)/dt| =1$ and $\mathrm{Re} = 1000$. As the plate's motion is purely normal to itself, pressure effects dominate, and we might expect that viscous effects can be neglected. Therefore, the oscillating plate represents an ideal situation for the inviscid model to accurately model direct viscous simulation. 
\newline

\begin{figure}[H]
    \centering
    \includegraphics[width=0.7\linewidth, trim=2cm 3cm 2cm 2cm, clip]{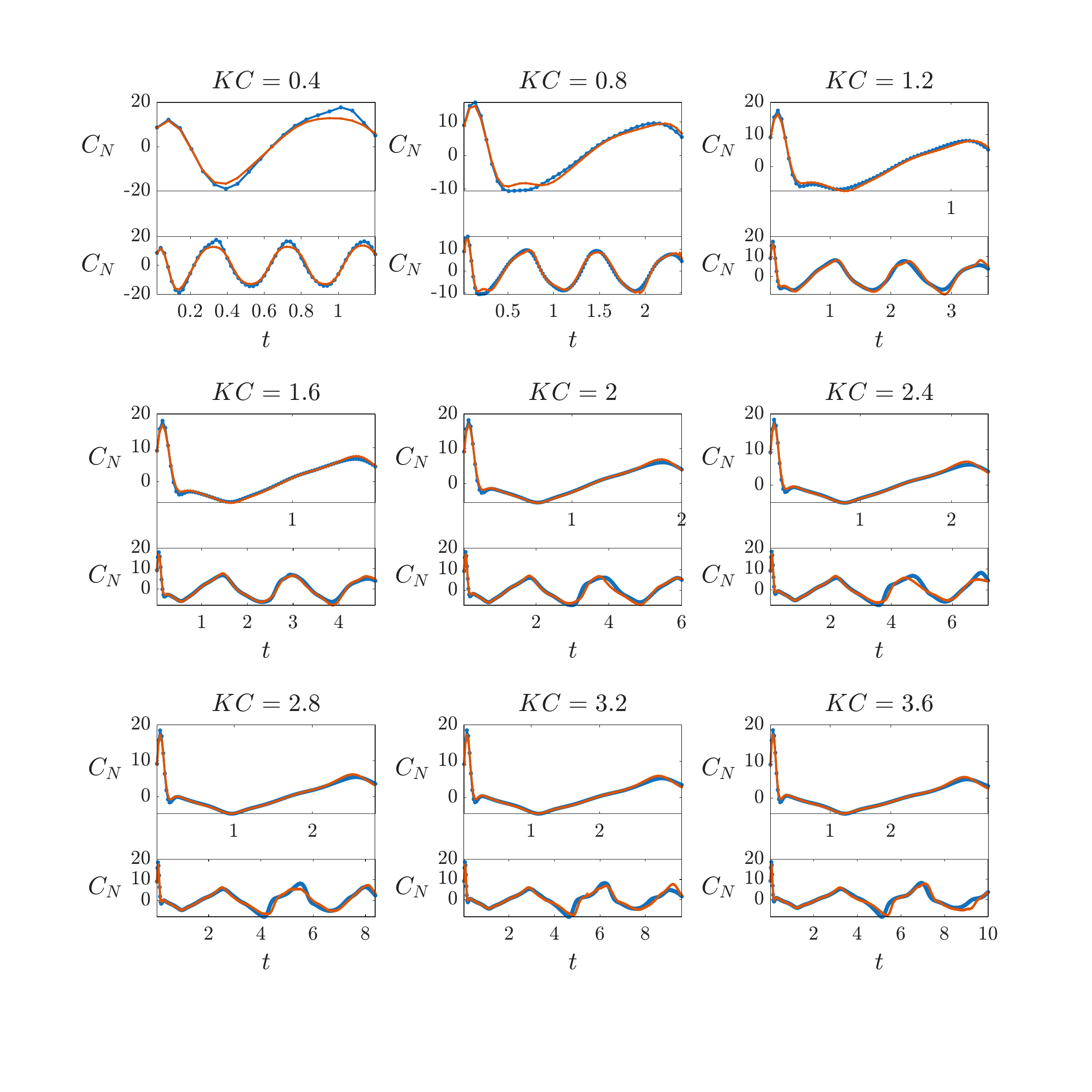}
    \caption{The net normal force up to time $t = \textrm{KC}$ (top) and time $t = 3\textrm{KC}$ (bottom) for both the viscous (blue) and inviscid (orange) models.}
    \label{oscillation force comparison figure}
\end{figure}

\begin{figure}[H]
    \centering
    \includegraphics[width=0.98\linewidth, trim=1.5cm 1cm 0cm 0cm, clip]{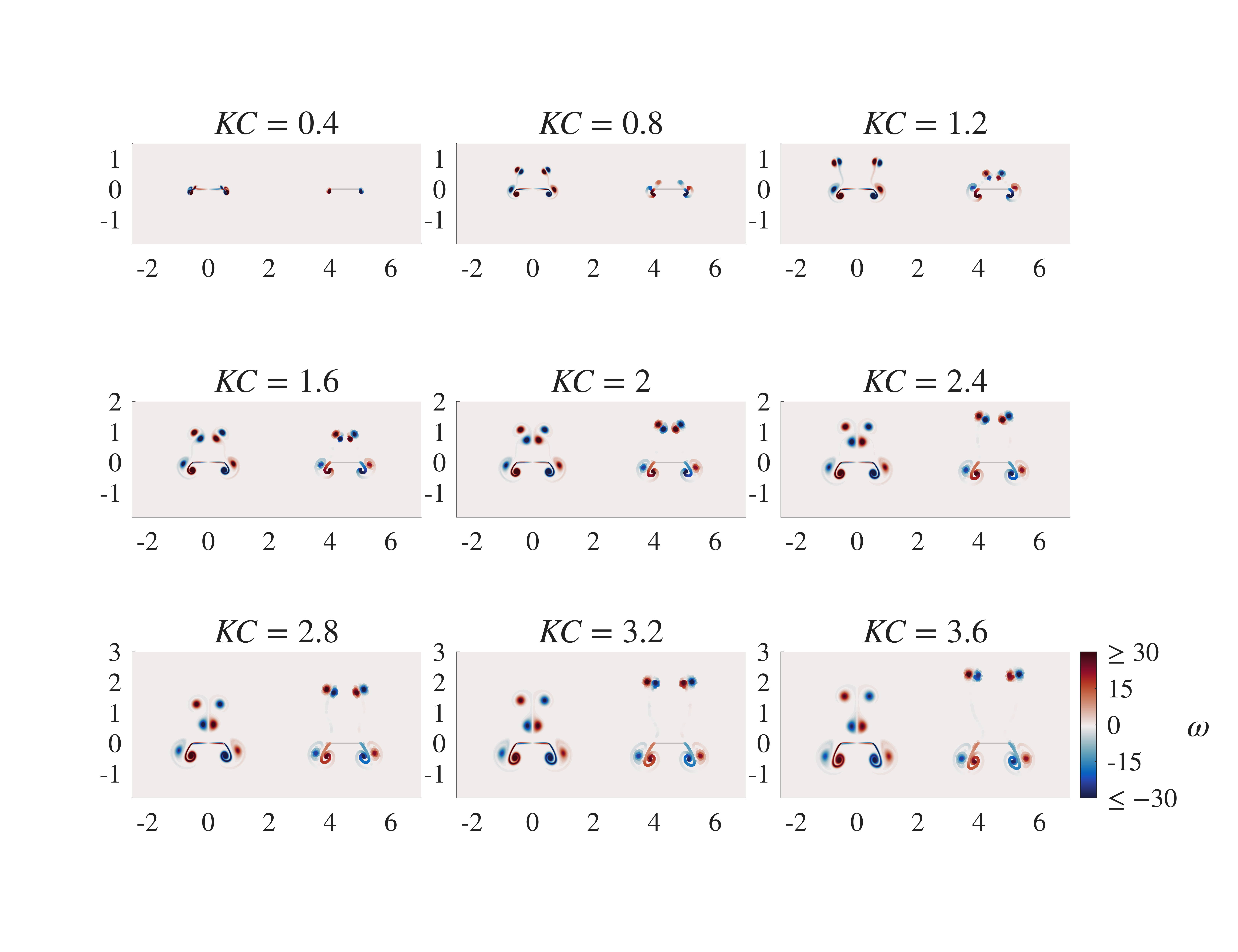}
    \caption{Vorticity $\omega$ at $t = \textrm{KC}$ for an oscillating plate with $\textrm{KC} = 0.4,\ \ldots, 3.6$ in inviscid (right) and viscous (left) models.}
    \label{oscillation vorticity comparison figure}
\end{figure}

In figure \ref{oscillation force comparison figure} we show a $3\times3$ grid of plots. Each plot corresponds to a different value of KC and is divided into a top and bottom panel. 
In the top and bottom panels we plot the net normal force $C_N(t)$ for both viscous (blue) and inviscid (orange) models up to $t = \textrm{KC}$ and $t = 3\textrm{KC}$ respectively. In figure \ref{oscillation vorticity comparison figure} we show a $3\times3$ grid of vorticity plots at the corresponding KC values. In each plot, the viscous flapping plate is placed on the unit interval $[-1/2,\ 1/2]\times\{0\}$ while the inviscid plate is placed 9/2 units to the right at $[4,\ 5]\times\{0\}$.
\newline

\begin{figure}[H]
    \centering
    \includegraphics[width=1\linewidth, trim=0cm 0cm 0cm 0cm, clip]{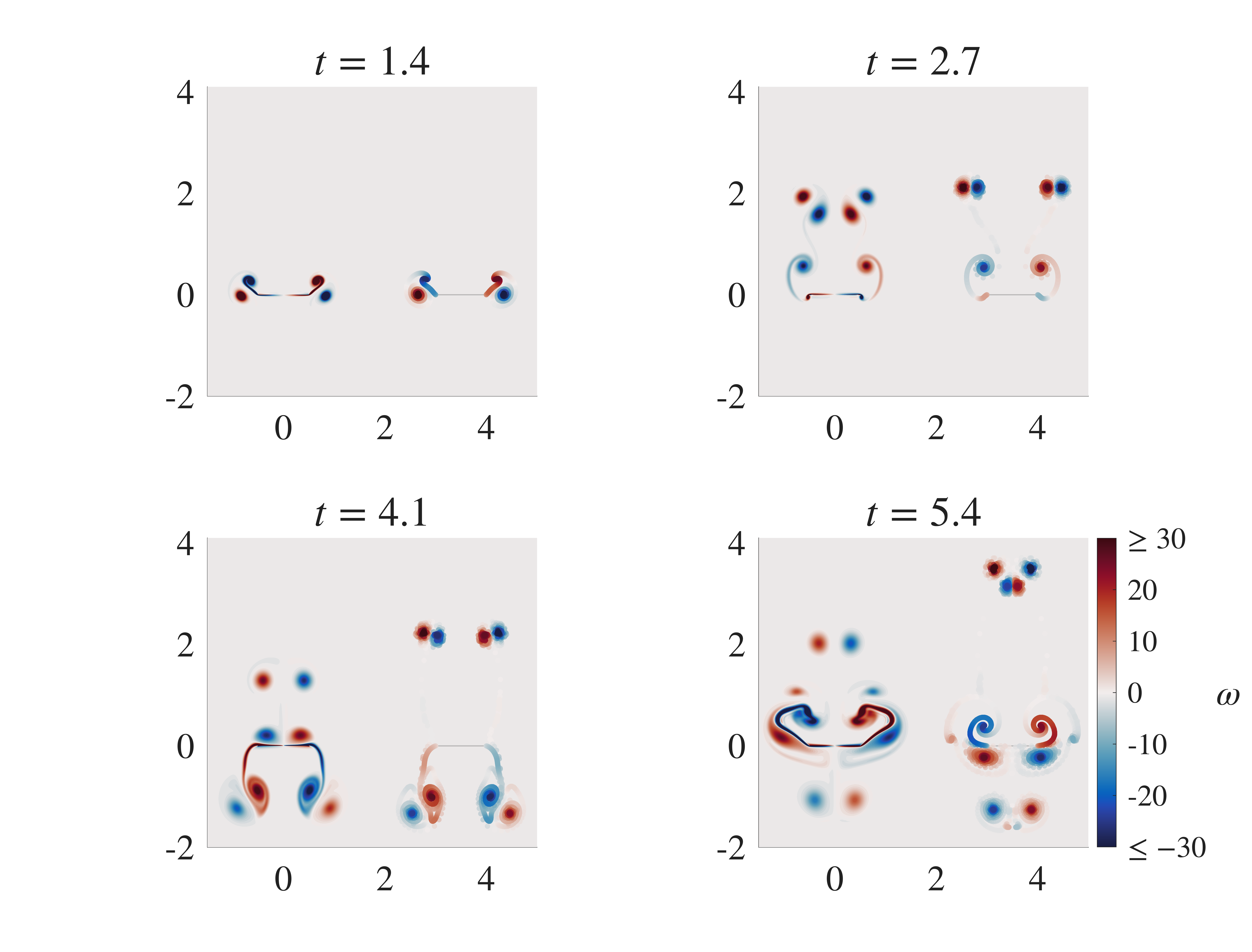}
    \caption{Vorticity fields at times $t = 1.4,2.7,4.1,$ and 5.4 for an oscillating plate with $\textrm{KC} = 3.6$ in inviscid (right) and viscous (left) models.}
    \label{oscillation vorticity example figure}
\end{figure}

Figure \ref{oscillation force comparison figure} shows good agreement between the computed normal force in both the viscous and inviscid models. Figure \ref{oscillation vorticity comparison figure} shows relatively worse agreement in the global vorticity fields. One reason for this discrepancy is the chaotic nature of 2D vortex dynamics \cite{aref1983integrable} which amplifies small discrepancies exponentially over time. Small differences between the models gradually introduce errors in the global vorticity fields that are later amplified by chaos. In general however, the most recently shed vorticity near the edges of the plate in both viscous and inviscid models match up well.  This can be seen in the panels corresponding to $\textrm{KC}$ = 1.6--3.6, and to a smaller extent the panels corresponding to $\textrm{KC}$ = 0.4--1.2. Figure \ref{oscillation vorticity example figure} shows snapshots of the vorticity fields of an oscillating plate with $\textrm{KC} = 3.6$ at times $t = 1.4, 2.7,4.1,5.4$ (i.e.~over 1.5 flapping periods). At early times the vorticity fields match up well, but by $t = 4.1$ these small differences have amplified, and the positions of the colliding dipoles above the plate are quite different. Examining $KC$ = 3.6 in figure \ref{oscillation force comparison figure}, we see that $C_N(t)$ is similar in the two models but with a phase shift that increases with time. It would seem that the aerodynamic forces on the plate are tightly coupled to the local vorticity field near the body of the plate and less sensitive to the global vorticity dynamics.  A similar observation was made in \cite{Koumoutsakos_Shiels_1996}; since the net normal force on the body is an integrated quantity, it may be less sensitive to some details of the far-field flow such as the sizes and locations of the vortex cores. 

\subsection{Heaving and pitching}
\label{heaving and pitching subsection}
\subsubsection{Pure heaving and pitching}

We now consider heaving and pitching motions of a plate in a uniform flow parallel to the plate's time-averaged orientation, a model of animal locomotion \cite{lighthill1969hydromechanics}. By introducing motion parallel (as opposed to perpendicular) to the plate, one expects the effect of viscosity to increase, leading to relatively worse agreement. For a heaving plate, we have 
\begin{align}
    \boldsymbol{X}_G(t) &= -[0,\frac{\text{St KC}}{2}\sin(\frac{2\pi}{\textrm{KC}} t)]^T  -[t,0]^T, \\
    \beta(t) &= 0.
\end{align}

Similarly, for a plate  pitching about its leading edge we have
\begin{align}
    \boldsymbol{X}_G(t) &= -[1/2-\cos(\beta(t))/2,-\sin(\beta(t))/2]^T - [t,0]^T,  \\
    \beta(t) &= -\arcsin\left(\frac{\text{St KC}}{2}\right)\sin(\frac{2\pi}{\textrm{KC}} t).
\end{align}

\begin{figure}[H]
    \centering
    \includegraphics[width=1\linewidth, trim = 4cm 4cm 5cm 1cm,clip]{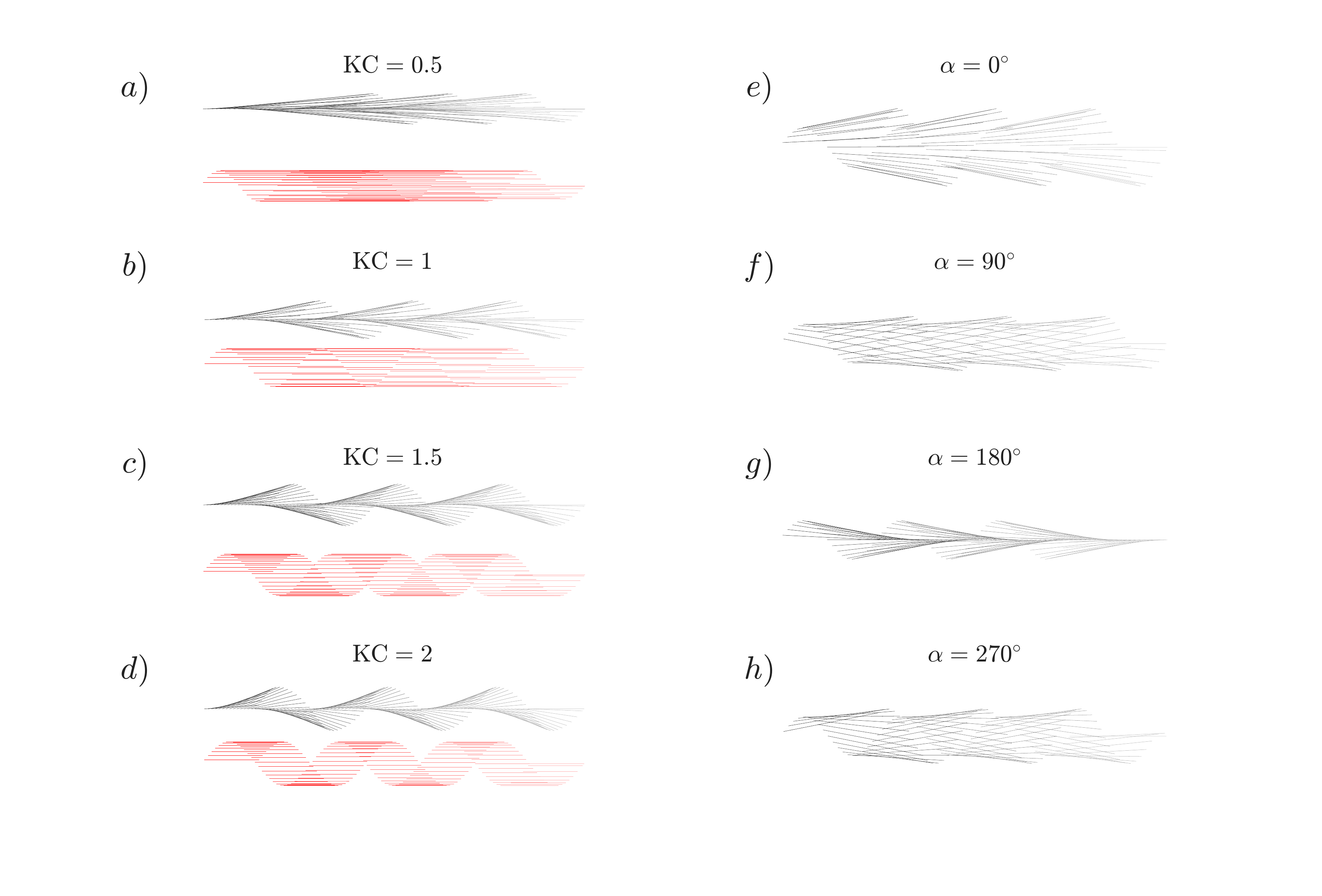}
    \caption{(a--d) Snapshots of a plate undergoing pitching (black) and heaving (red)  with St = 0.4 and $\textrm{KC} =0.5,1,1.5,2$ respectively. (e--g) Snapshots of a plate undergoing a mixed pitching-heaving motion with $\textrm{St} = 0.4$, $\textrm{KC} =1$ and  phase offsets $\alpha = 0^\circ,90^\circ,180^\circ,270^\circ$ respectively. The shading of the plates ranges from light to dark as time proceeds.}
    \label{fig:mixed pitchheave explanation figure}
\end{figure}

Note that the inclusion of the $[-t,0]^T$ term is equivalent to introducing a background flow $[1,0]^T$. Here we consider Strouhal numbers $\text{St}$ = 0.2 and 0.4 and $\textrm{KC}$ = 0.5, 1, 1.5, and 2 for both pitching and heaving foils.  In figure \ref{fig:mixed pitchheave explanation figure} we show snapshots of a plate undergoing these motions. Panels a--d show snapshots of a plate undergoing pitching (black) and heaving (red) at $\textrm{St} = 0.4$ for $\textrm{KC}=0.5,1,1.5,2$ respectively. Panels e--h show the same for a plate undergoing a mixed pitching-heaving motion with St = 0.4, KC = 1 and  phase offsets $\alpha = 0^\circ,90^\circ,180^\circ,270^\circ$ respectively. In the subsequent section \ref{mixed pitchheave section} we discuss the mixed pitching-heaving motions.

\begin{figure}[H]
    \centering
    \includegraphics[width=1\linewidth]{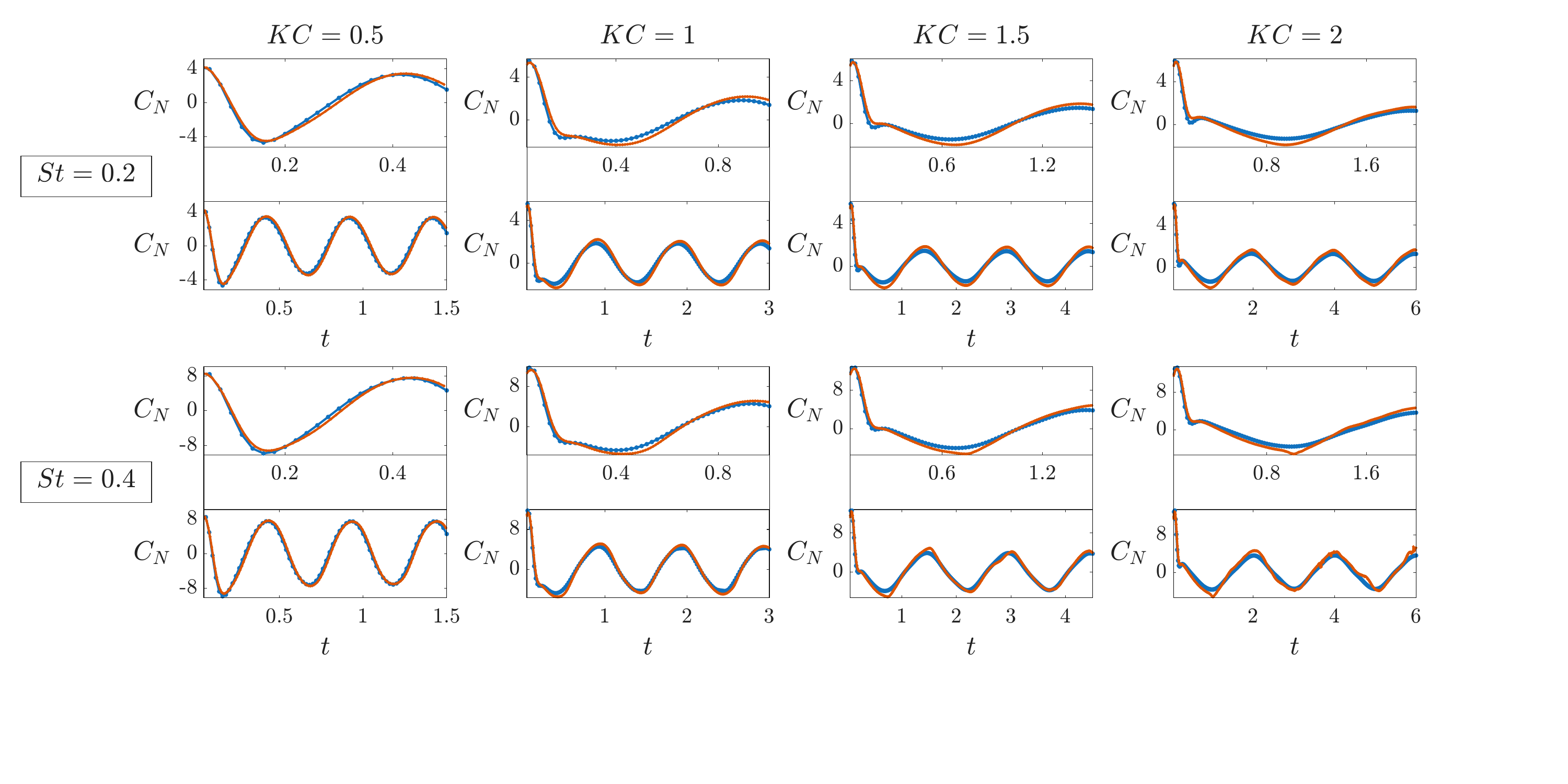}
    \caption{Net normal force on a pitching plate in viscous (blue) and inviscid (orange) flow, at four KC values. The top row shows results for $\textrm{St} = 0.2$ while the bottom row shows results for $\textrm{St} = 0.4$. Each plot is divided into a top and bottom panel, showing the net normal force $C_N$ up to time $t=\textrm{KC}$ and $t = 3\textrm{KC}$ respectively.}
    \label{fig:pitchheave pitching force comparison}
\end{figure}

\begin{figure}[H]
    \centering
\includegraphics[width=1\linewidth, trim = 4cm 0.5cm 3cm 2cm, clip]{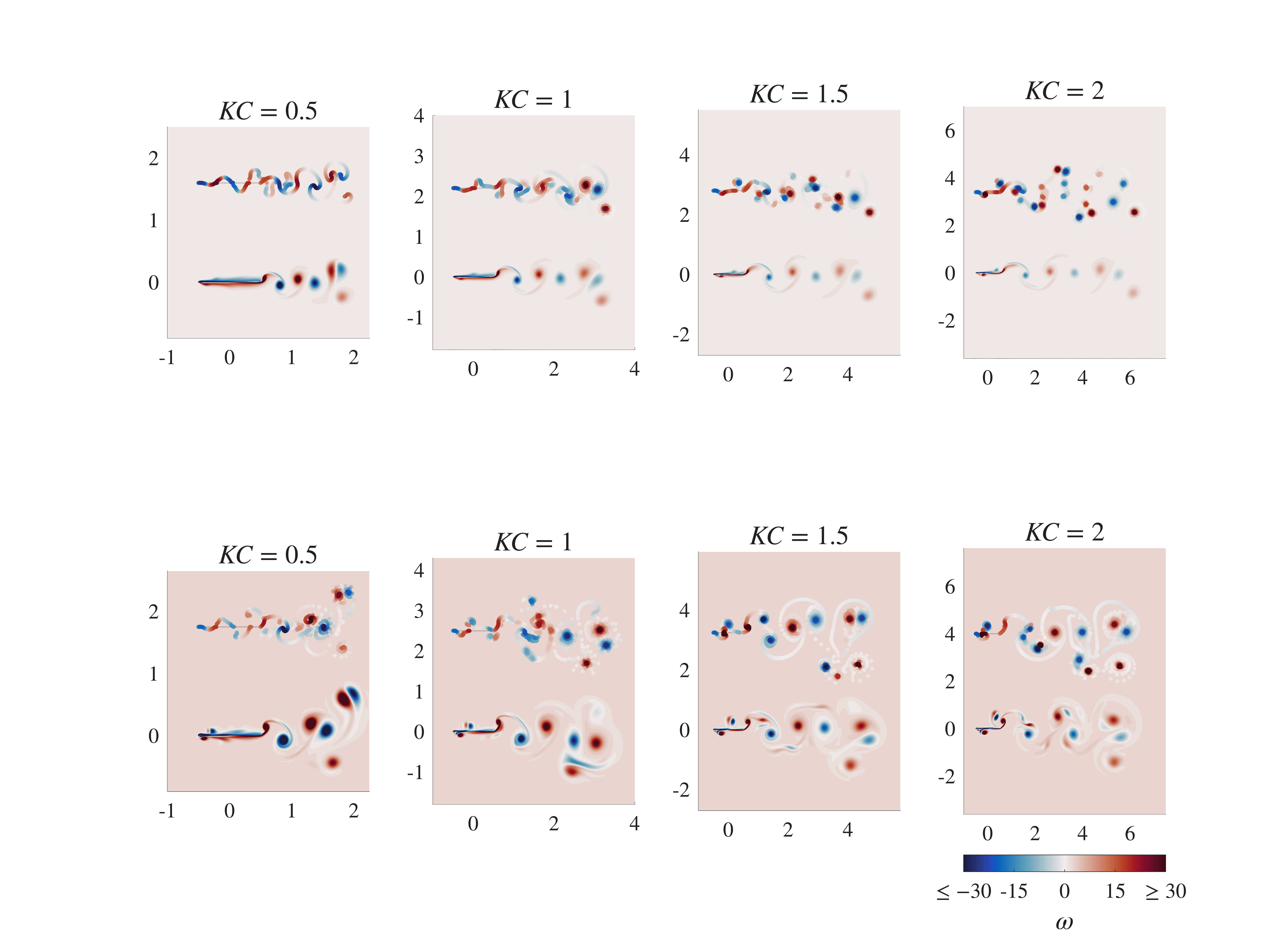}
    \caption{Vorticity $\omega$ at time $t = 3\textrm{KC}$ for a pitching plate in both viscous (bottom) and inviscid (top) models, at four KC values. The top row shows results for $\textrm{St} = 0.2$ while the bottom row shows results for $\textrm{St} = 0.4$.}
    \label{fig:pitchheave pitching vorticity comparison}
\end{figure}

\begin{figure}[H]
    \centering
    \includegraphics[width=1\linewidth]{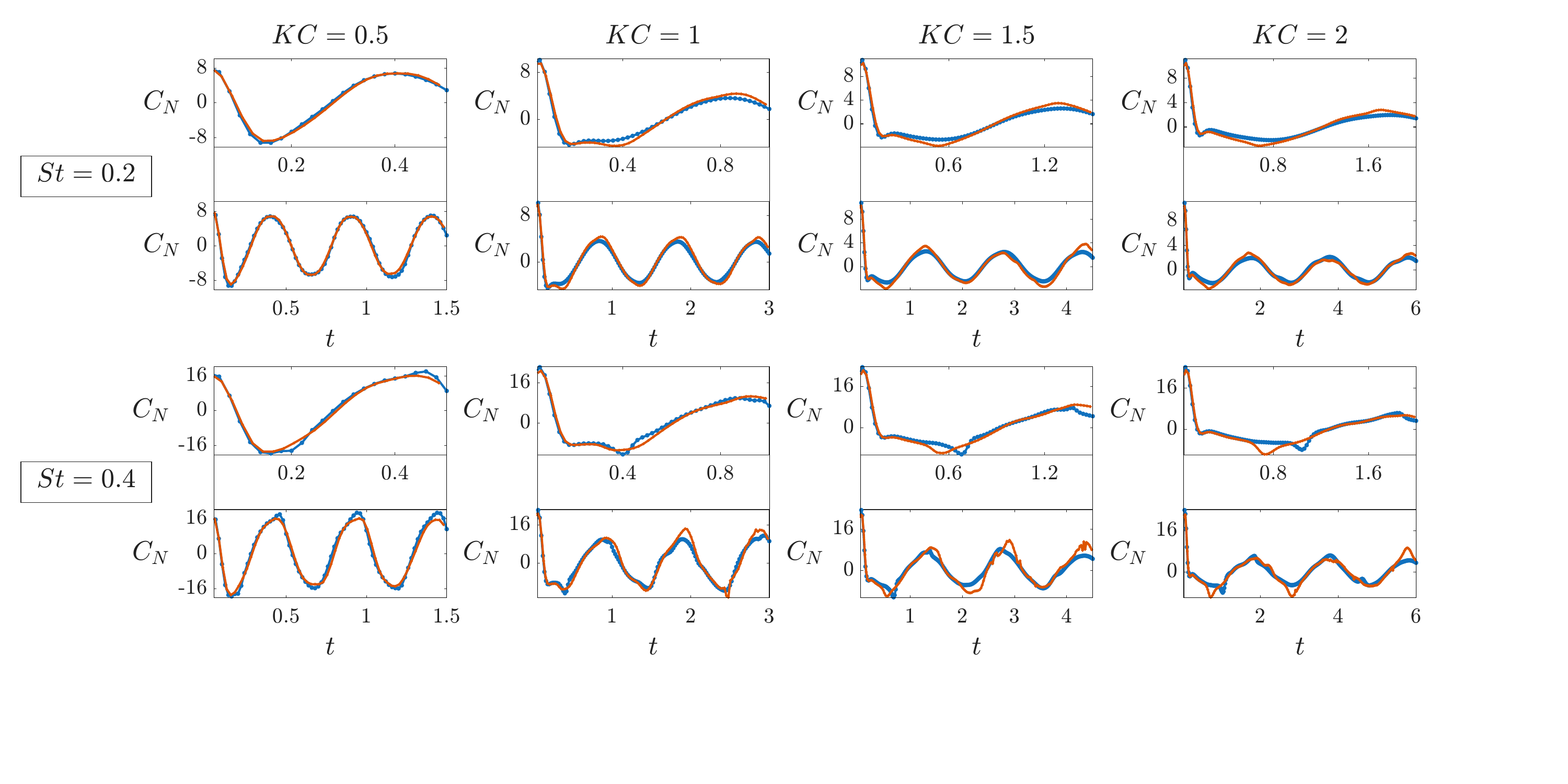}
    \caption{Net normal force on a heaving plate in viscous (blue) and inviscid (orange) flow, at four KC values. The top row shows results for $\textrm{St} = 0.2$ while the bottom row shows results for $\textrm{St} = 0.4$. Each panel is divided into a top and bottom subpanel, showing the normal force $C_N$ up to time  $t=\textrm{KC}$ and $t = 3\textrm{KC}$ respectively.}
    \label{fig:pitchheave heaving force comparison}
\end{figure}

\begin{figure}[H]
    \centering
\includegraphics[width=1\linewidth, trim = 4cm 0.1cm 3cm 2cm, clip]{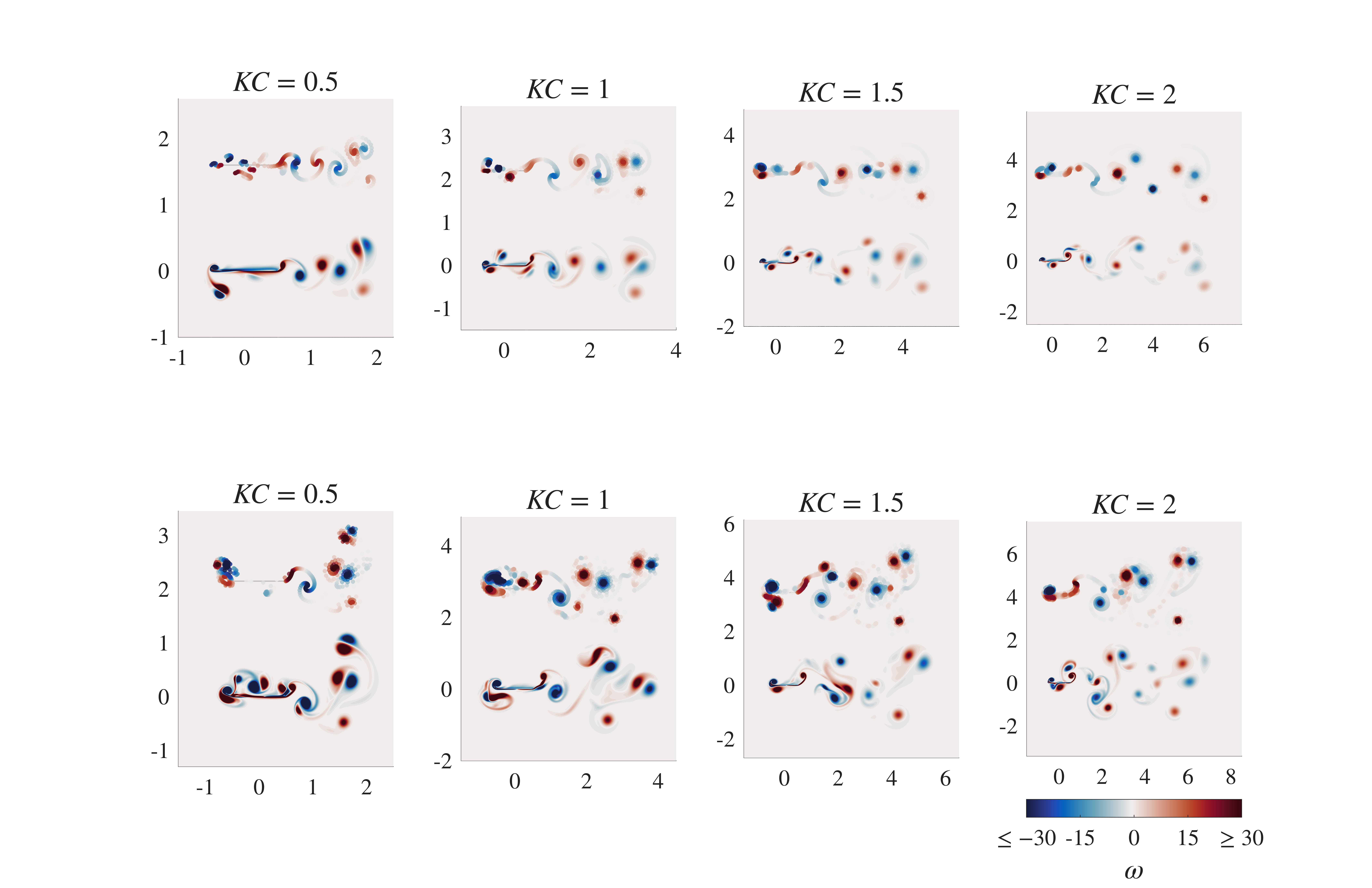}
    \caption{Vorticity $\omega$ at time $t = 3\textrm{KC}$ for a heaving plate in both viscous (bottom) and inviscid (top) models, at four KC values. The top row shows results for $\textrm{St} = 0.2$ while the bottom row shows results for $\textrm{St} = 0.4$.}
    \label{fig:pitchheave heaving vorticity comparison}
\end{figure}

As in the previous section, KC is the dimensionless period. Increasing KC thus decreases the dimensionless frequency, which in turn increases the strength of the resulting leading-edge vortex \cite{rival2014characteristic, anderson1998oscillating}. By contrast, experiments suggest that varying the Strouhal number has less of an effect on leading-edge vortex formation \cite{rival2014characteristic}.  Figures \ref{fig:pitchheave pitching force comparison} and \ref{fig:pitchheave heaving force comparison} show $C_N(t)$ for pitching and heaving plates respectively. In both figures the plots are presented in a $2\times4$ grid, each column corresponding to a different KC, increasing left to right, and the top and bottom rows corresponding to $\textrm{St} = 0.2$ and 0.4 respectively. As before, each panel is divided into a top and bottom subpanel showing the net normal force $C_N$ for both viscous (blue) and inviscid (orange) up to times $t = \textrm{KC}$ and $t =3\textrm{KC}$ respectively (i.e.~1 and 3 flapping periods). Figures \ref{fig:pitchheave pitching vorticity comparison} and \ref{fig:pitchheave heaving vorticity comparison} show the corresponding vorticity fields for pitching and heaving plates respectively, in the viscous and inviscid flows at time $t = \textrm{KC}$. In each panel, the viscous and inviscid plates are placed at $[-1/2,1/2]\times\{0\}$ and $[-1/2,1/2]\times\{1 + 1.2\ \textrm{KC}\}$ respectively.
\newline 

In the viscous simulations, one often observes vortices rolling up at the leading edge, which later form small vortex dipoles that travel along the surface of the plate before being released at the trailing edge. This is readily apparent in the bottom row of figure \ref{fig:pitchheave pitching vorticity comparison}, at $\textrm{St}=0.4$. Indeed, in all four panels one observes a recently released dipole above the plate, and the formation of a small leading-edge vortex below the plate in both the inviscid and viscous models. This form of vortex generation is neglected in vortex-sheet models without leading-edge shedding and can significantly affect the resulting forces \cite{PanXiaoLE}. An investigation similar to the present study is done for a pitching plate for the single case with $\textrm{St} = 0.4$ and $\textrm{KC} = 0.5$ in \cite{SimulatingVortexFlapping} for an inviscid model \textit{without} leading-edge shedding and obtains relatively worse agreement in the computed forces than the present model does with continuous leading-edge shedding. 
\newline

At the same time, however, the absence of viscous diffusion in the inviscid model means that these leading-edge vortices do not dissipate. In the viscous model, by the time the leading-edge vortices have traveled near the trailing edge of the plate, they have almost completely dissipated, and hence do not affect the wake topology. This is particularly apparent in the viscous model when KC is small (and hence the leading-edge vortex is weak). This can be clearly observed in the panels corresponding to St = 0.2 at KC = 0.5 and 1 in figure \ref{fig:pitchheave pitching vorticity comparison} for the pitching plate, where the viscous wake topology is characterized almost completely by the vorticity shed at the trailing edge. 
\newline

As KC increases, the vortex sheets transition from appearing sheet-like to appearing as discrete vortices in figures \ref{fig:pitchheave pitching vorticity comparison} and \ref{fig:pitchheave heaving vorticity comparison}. This is because the vorticity snapshots are taken at time 3KC, so when KC is larger, the vortex sheets will have had more time to  roll up (due to the Kelvin-Helmholtz instability). More specifically, note that the blob-regularized Biot-Savart kernel has symmetry $K_\delta(\boldsymbol{x}) = \frac{1}{\textrm{KC}}K_{\delta/\textrm{KC}}(\boldsymbol{x}/\textrm{KC})$. If we then consider the rescaled time, space and regularization parameters $\tilde{t} =t/\textrm{KC}$, $\tilde{\boldsymbol{x}} = \boldsymbol{x}/\textrm{KC}$, and $\tilde{\delta} = \delta/\textrm{KC}$, the motion of the heaving plate can be written as 

\begin{align}
    \tilde{\boldsymbol{X}}_G(t) &= -[0,\frac{\text{St}}{2}\sin(2\pi\tilde{t})]^T -[\tilde{t},0]^T, \\
    \tilde{\beta}(t) &= 0.
\end{align}

Notice the motion is now independent of $\textrm{KC}$. Hence for fixed St,  the effect of changing KC on the resulting dynamics is equivalent to scaling the plate length by a factor of $\frac{1}{\textrm{KC}}$ and evolving it with the apparent regularization parameter $\tilde{\delta} = \frac{\delta}{\textrm{KC}}$. The pitching case is similar. Hence heuristically, for fixed $\delta$, shrinking KC increases the apparent regularization of the sheets $\frac{\delta}{\textrm{KC}}$. It is well known that increasing the regularization parameter delays vortex sheet roll-up, causing it to appear more sheet-like \cite{Krasny_1986, loo2025falling}. On the other hand, with larger KC, $\delta$ is smaller relative to KC, making the vortex sheet more point-like. In general, tuning $\delta$ can make the inviscid and viscous vortex wakes match more closely.
\newline

The vorticity fields for the heaving plate show comparatively worse agreement between inviscid and viscous models. Indeed, for a heaving plate, the leading edge moves in both the $x$ and $y$ directions as opposed to the pitching plate, whose leading edge moves only in the $x$ direction. The leading-edge vortex dynamics for the heaving plate is much more complicated. As the plate reverses direction, strong dipoles are released from the leading edge (e.g.~KC $= 0.5$ and St $= 0.2$ in figure \ref{fig:pitchheave heaving vorticity comparison}). This leading-edge vortex often collides with the body of the plate leading to secondary vortex formation through complex boundary layer interactions that inviscid model does not represent. As the heaving-plate flow is more sensitive to the leading-edge vortex dynamics, the inviscid model exhibits generally lower accuracy than it does for pitching. In addition, the heaving-plate flow appears more sensitive to the Strouhal number. In figures  \ref{fig:pitchheave heaving force comparison} and \ref{fig:pitchheave heaving vorticity comparison}, the inviscid model exhibits more accuracy at $\textrm{St} =0.2$ (top row) than at $\textrm{St}=0.4$. When the Strouhal number is lower, the leading-edge vorticity is smaller in magnitude, and the dynamics of the dipoles released at the leading edge is less chaotic. By contrast, the Strouhal number of the pitching plate has a weaker effect on the accuracy of the inviscid model---good accuracy is attained at both Strouhal numbers.

\subsubsection{Mixed heaving and pitching}
\label{mixed pitchheave section}
We now superpose the pure heaving and pitching motions we have introduced, for the single $\textrm{KC} = 1$, but for four different phase differences between heaving and pitching given by $\alpha = 0^\circ, 90^\circ, 180^\circ,$ and $270^\circ$. That is, we set

\begin{equation}
    \boldsymbol{X}_G(t) = -[t,\frac{\text{St KC}}{2}\sin(\frac{2\pi}{\text{KC}} t)]^T - [1/2-\cos(\beta(t))/2,-\sin(\beta(t))/2]^T
\end{equation}
and
\begin{equation}
    \beta(t) = -\arcsin(\frac{\text{St KC}}{2})\sin(\frac{2\pi}{\text{KC}} t + \alpha)
\end{equation} 

The motions are shown in figure \ref{fig:mixed pitchheave explanation figure}e--g. $\alpha = 0^\circ$ corresponds to heaving in phase with pitching, while $\alpha = 180^\circ$ is anti-phase. $\alpha = 90^\circ$ and $270^\circ$ are one-quarter and three-quarters of a period out of phase. In the latter motion the plate moves more tangentially to its path at the pitching axis, decreasing resistance in the normal direction. This case is the closest to the motions of the relatively-rigid crescent or ``lunate" tail fins of swimming tuna and mackerel, and has been studied by many authors including Lighthill, who termed the ratio of the plate slope to the path slope at the pitching axis the ``proportional feathering" parameter \cite{lighthill1969hydromechanics,chopra1974hydromechanics}, referring to the ``feathering" of aircraft wings. 
\newline

\begin{figure}[H]
    \centering
    \includegraphics[width=0.9\linewidth, trim = 0cm 4cm 0cm 0cm,clip]{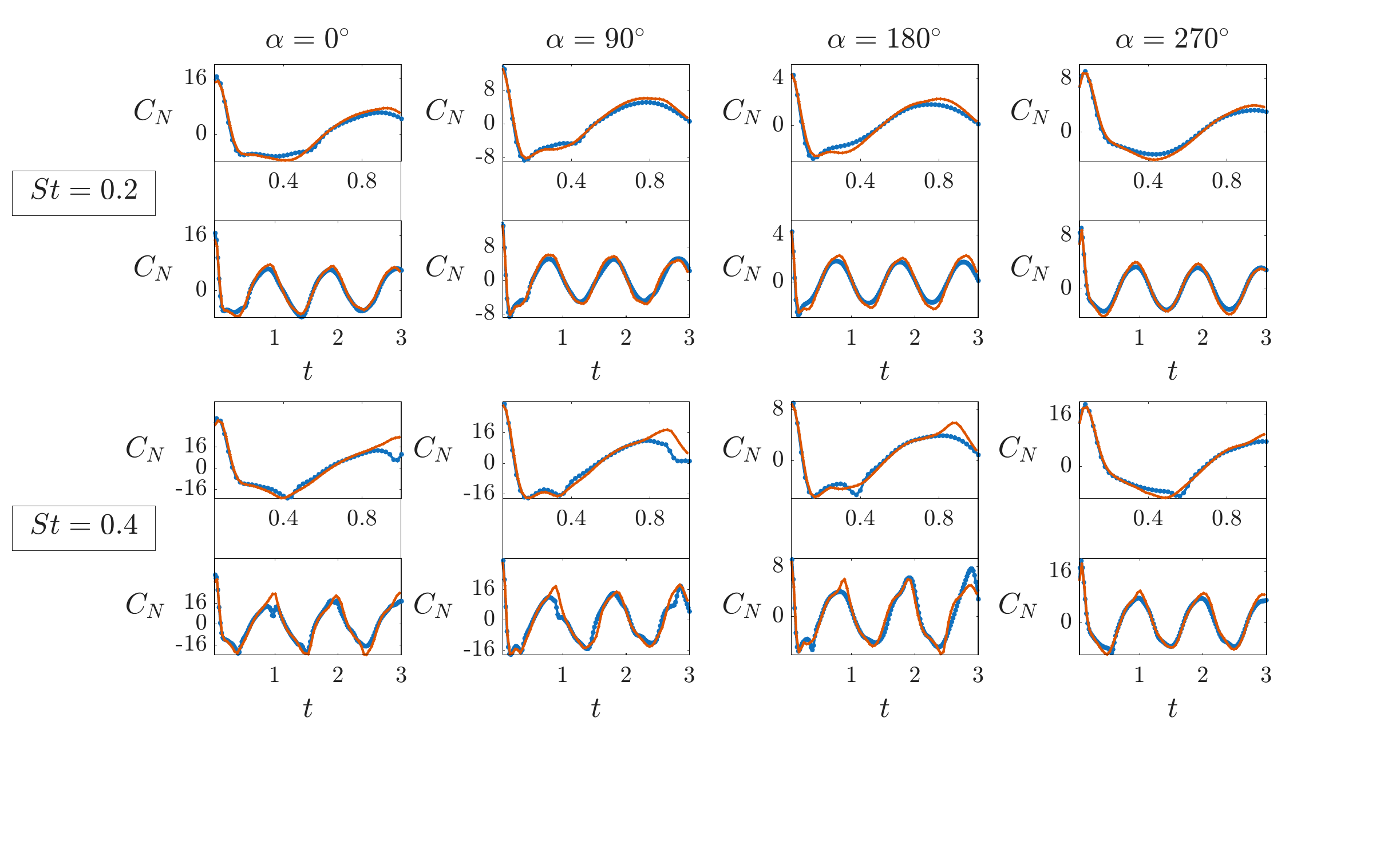}
    \caption{Net normal force on a heaving and pitching plate in viscous (blue) and inviscid (orange) flow, with $\textrm{KC} = 1$ and $\alpha$ varying across the columns. The top row shows results for $\textrm{St} = 0.2$ while the bottom row shows results for $\textrm{St} = 0.4$. Each plot is divided into a top and bottom panel, showing the net normal force $C_N$ up to time $t=$ KC and $t = 3$KC respectively.}
    \label{fig:pitchheave phaseshift force comparison KC1}
\end{figure}

In figure \ref{fig:pitchheave phaseshift force comparison KC1} 
we present the net normal forces for these motions with a $2\times4$ grid of panels. From left to right, the columns correspond to $\alpha$ = $0^\circ ,90^\circ, 180^\circ,$ and $270^\circ$. Each panel consists of a top and bottom subpanel showing the net normal force $C_N$ for the viscous (blue) and inviscid (orange) simulations up to times $t = \textrm{KC}$ and $t = 3\textrm{KC}$ respectively. 
The top row corresponds to $\textrm{St} = 0.2$ and the bottom row correspond to $\textrm{St} = 0.4$. The forces are largest when $\alpha = 0^\circ$ and smallest when
$\alpha = 180^\circ$, presumably because these are the cases with largest and smallest normal motions, as shown in figure \ref{fig:mixed pitchheave explanation figure}e--g.
\newline

One might expect the inviscid model to be least accurate at $\alpha = 270^\circ$ since the plate motion is more tangential in this case. Viscous effects might then dominate the flow field causing greater discrepancy in the computed forces. From figure \ref{fig:pitchheave phaseshift force comparison KC1} however, this does not appear to be the case. Indeed, the error in computed forces across $\alpha$ is mostly uniform.
\newline

It is worth noting that in the viscous model, vorticity is generated at the surface of the plate as a result of the no-slip condition, which is a purely viscous boundary condition. In the inviscid model however, vorticity is instead generated to enforce the no-penetration condition on the surface of the plate together with Kelvin's theorem and the Kutta conditions. Despite apparently different sources of vorticity and hence force generation, both appear to induce similar aerodynamic forces.

\subsection{Pitch-up}
\label{pitch up subsection}
In the examples considered thus far, the leading-edge vortex is relatively weak. We now turn to motions in which the leading-edge vortex is much stronger, focusing on the pitch-up motion, a configuration studied extensively in the literature \cite{eldredge2009computational, eldredge2010high}. Here we examine a comprehensive collection of pitch-up motions with a fixed angle of attack, $45^\circ$. We take
\begin{align}
    \boldsymbol{X}_G(t) &= [t,0]^T + (I  - R_{\beta(t)})[1/2 - X_p, 0]^T \\ 
     \beta(t) &=\alpha_0 \frac{G(t)}{\max_{0\leq \tau \leq t_4}{G}(\tau)}
\end{align}
where 
\begin{equation}
G(t) = \ln\bigg(\frac{\cosh(a_s(t -t_1))\cosh(a_s(t - t_4))}{\cosh(a_s(t-t_2))\cosh(a_s(t-t_3))}\bigg).
\end{equation}

Here, $\alpha_0 = \frac{\pi}{4},\ a_s = 11,\ t_1 = 1,\ t_2 = t_1 + \frac{\alpha_0}{2K},\ t_3 = t_2 + 1.12, \ t_4 = t_3 + \frac{\alpha_0}{2K}$ as was done in \cite{eldredge2010high}, $I$ is the $2\times2$ identity matrix and $R_{\beta(t)} = \begin{bmatrix}
 \cos(\beta(t)) & -\sin(\beta(t)) \\
 \sin(\beta(t)) & \cos(\beta(t))
\end{bmatrix}$.
\newline

\begin{figure}[H]
    \centering
    \includegraphics[width=1\linewidth]{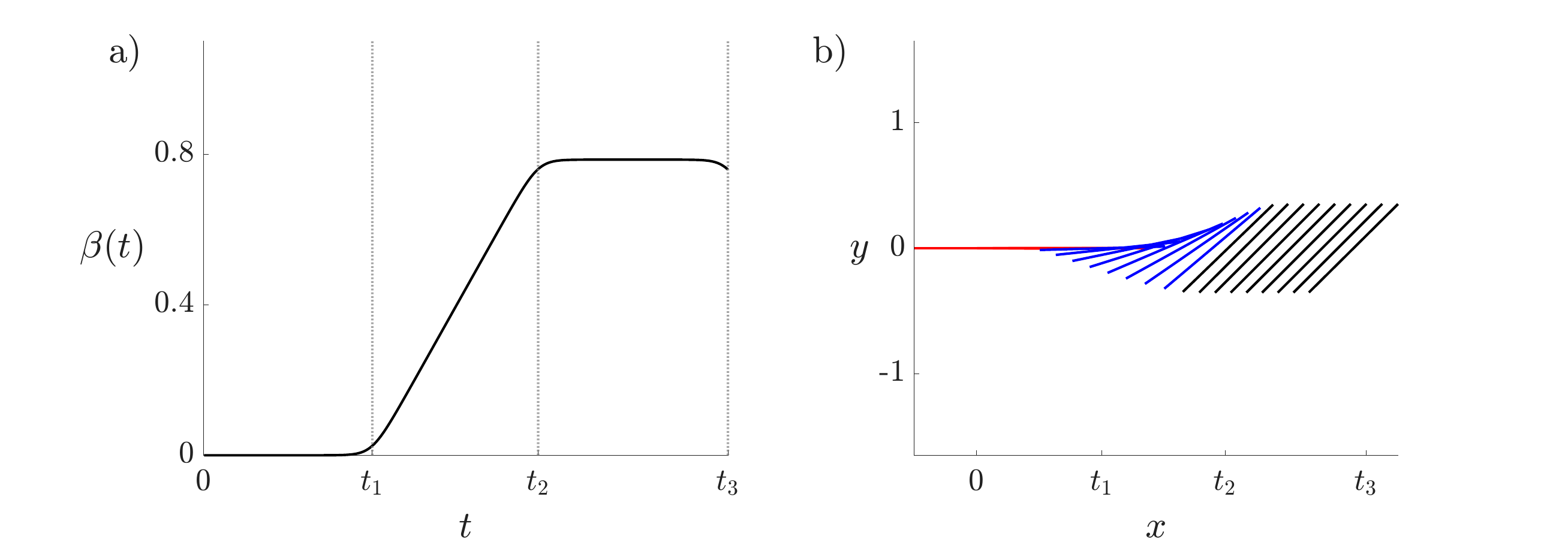}
    \caption{ (a) The orientation angle $\beta(t)$ of a plate undergoing a pitch-up motion (b) Snapshots of plate position while undergoing a pitch-up motion with pivot $X_p = 0.5$ in red, blue and black corresponding to the time intervals $[0,t_1], [t_1,t_2], [t_2,t_3]$ respectively.}
    \label{pitch up explanation figure}
\end{figure}

In plain language, the plate angle is zero from $t=0$ to $t_1$. From $t=t_1$ to $t_2$ the plate angle rises from $0$ to $\alpha_0$ approximately linearly (hence with angular velocity $\approx 2K$). From $t=t_2$ to $t_3$ the plate orientation remains constant at $\alpha_0$. In figure \ref{pitch up explanation figure}a we show how the orientation angle of the plate evolves. In panel b we show snapshots of the plate undergoing a pitch-up motion in red, blue and black, corresponding to the time intervals  $[0,t_1], [t_1,t_2], [t_2,t_3]$ respectively. We consider $K = 0.2,\ 0.4,\  \ldots, 1$ and pitch axis location $X_p = 0,\ 0.25,\ 0.5$. $X_p$ = 0 corresponds to pitching about the leading edge and $X_p$ = 0.5 corresponds to pitching about the plate center.

\begin{figure}[H]
    \centering
    \includegraphics[width=1\textwidth, trim=4cm 0cm 4cm 0cm, clip]{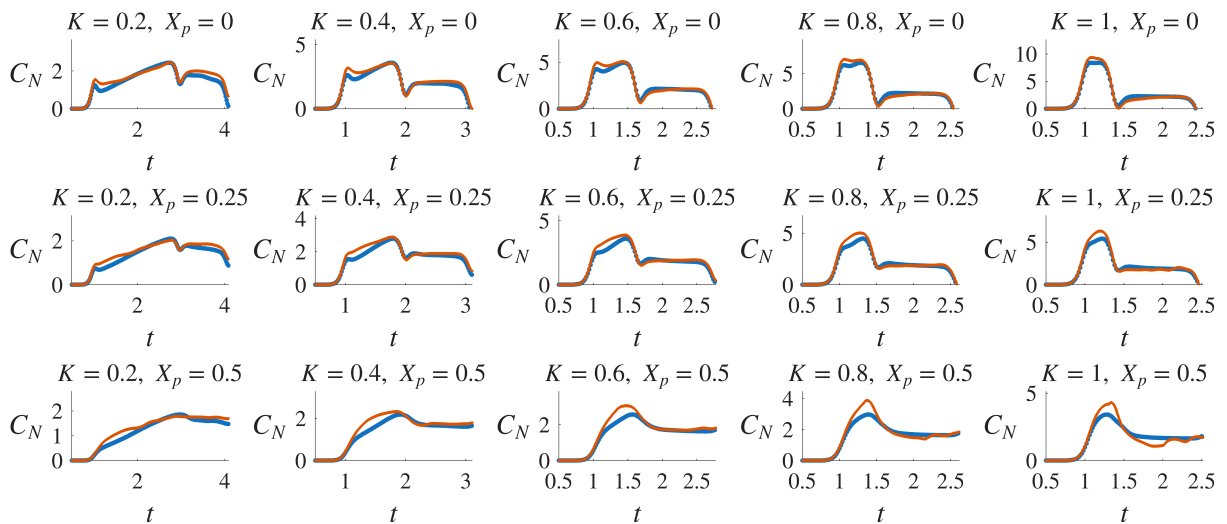}
    \caption{Net normal force during pitch-up up to time $t = t_3$ for both the viscous (blue) and inviscid (orange) models. The pitch rate $K$ increases from the left column to the right column, and the pitch axis location $X_p$ increases from the top row to the bottom row.}
    \label{force comparison pitch up}
\end{figure}

\begin{figure}[H]
    \centering
\includegraphics[width=1\textwidth, trim=2cm 0cm 2cm 0cm, clip]{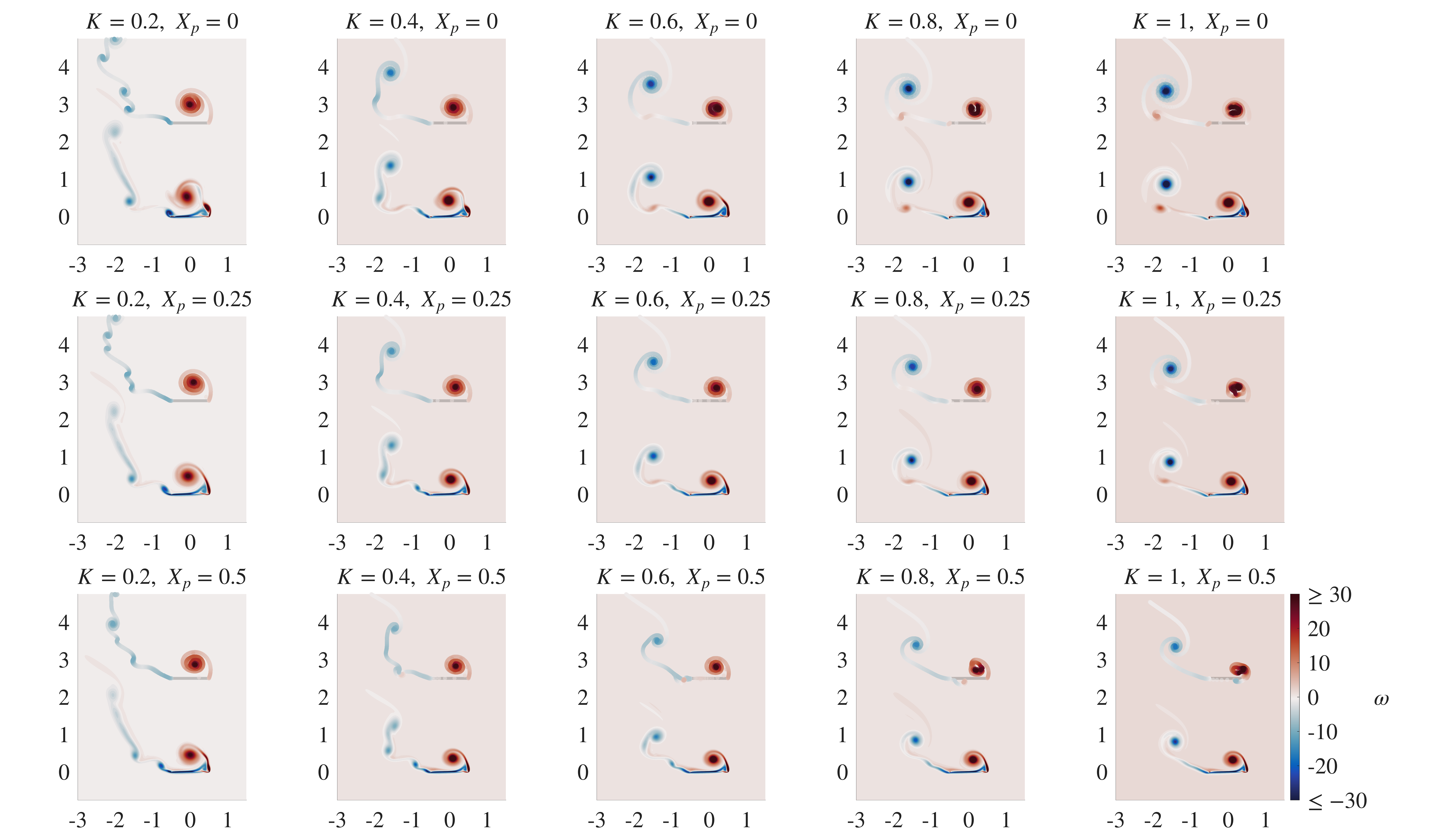}
    \caption{Vorticity $\omega$ at time $t = t_3$ for plate undergoing a pitch-up maneuver in both viscous (bottom) and inviscid (top) models.}
    \label{vortex comparison pitch up}
\end{figure}

Figure \ref{force comparison pitch up} shows a $3\times5$ grid of plots of the net normal force $C_N$ over time. Each plot corresponds to a different pair of values of $K$ and $X_p$, with the inviscid and viscous normal forces in orange and blue respectively. Figure \ref{vortex comparison pitch up} shows a $3\times5$ grid of the corresponding vorticity fields at the final times, with the viscous and inviscid plates placed at $[-1/2,1/2]\times\{0\}$ and $[-1/2,1/2]\times \{2.5\}$ respectively.
\newline 

The pitch-up motion is a rotary motion about a pitch axis on the plate. During the pitch-up (blue snapshots in figure \ref{pitch up explanation figure}b), the rotary motion contributes a large component of plate velocity perpendicular to the plate.
The pitch-up maneuver therefore appears to be a highly pressure-driven motion. Hence, much like the oscillating plate, we expect the inviscid model to accurately capture the resulting aerodynamic forces. Figure \ref{force comparison pitch up} shows that this generally holds. The error is mostly uniform among these plots, but is largest when $X_p = 0.5$ and $K \geq 0.4$. One reason for the discrepancy at these values of $(K,X_p)$ is that the local vorticity fields near the plate predicted by the viscous and inviscid models are somewhat different. This can be seen most clearly in the panel corresponding to $K = 0.8$ and $X_p = 0.5$ in figure \ref{vortex comparison pitch up}. Below the trailing edge of the top plate (inviscid) is a small dipole. When the plate pitches upward at $t = t_1$, for $X_p = 0.5$ and $K \geq 0.4$ a dipole is released below the plate. In the viscous models however, no dipole forms below the plate. This may be due to viscous effects at the leading edge that suppress the formation of a concentrated dipole at the leading edge through dissipation. At the larger Reynolds number $\mathrm{Re} = 2000$, dipoles do form at the leading edge for a plate undergoing a somewhat different pitching motion \cite{moubogha2017forces}.
\newline

\subsection{Uniform rotation}
\label{uniform rotation subsection}
\begin{figure}[H]
    \centering
\includegraphics[width=1\linewidth, trim = 2.5cm 1cm 3cm 0cm, clip]{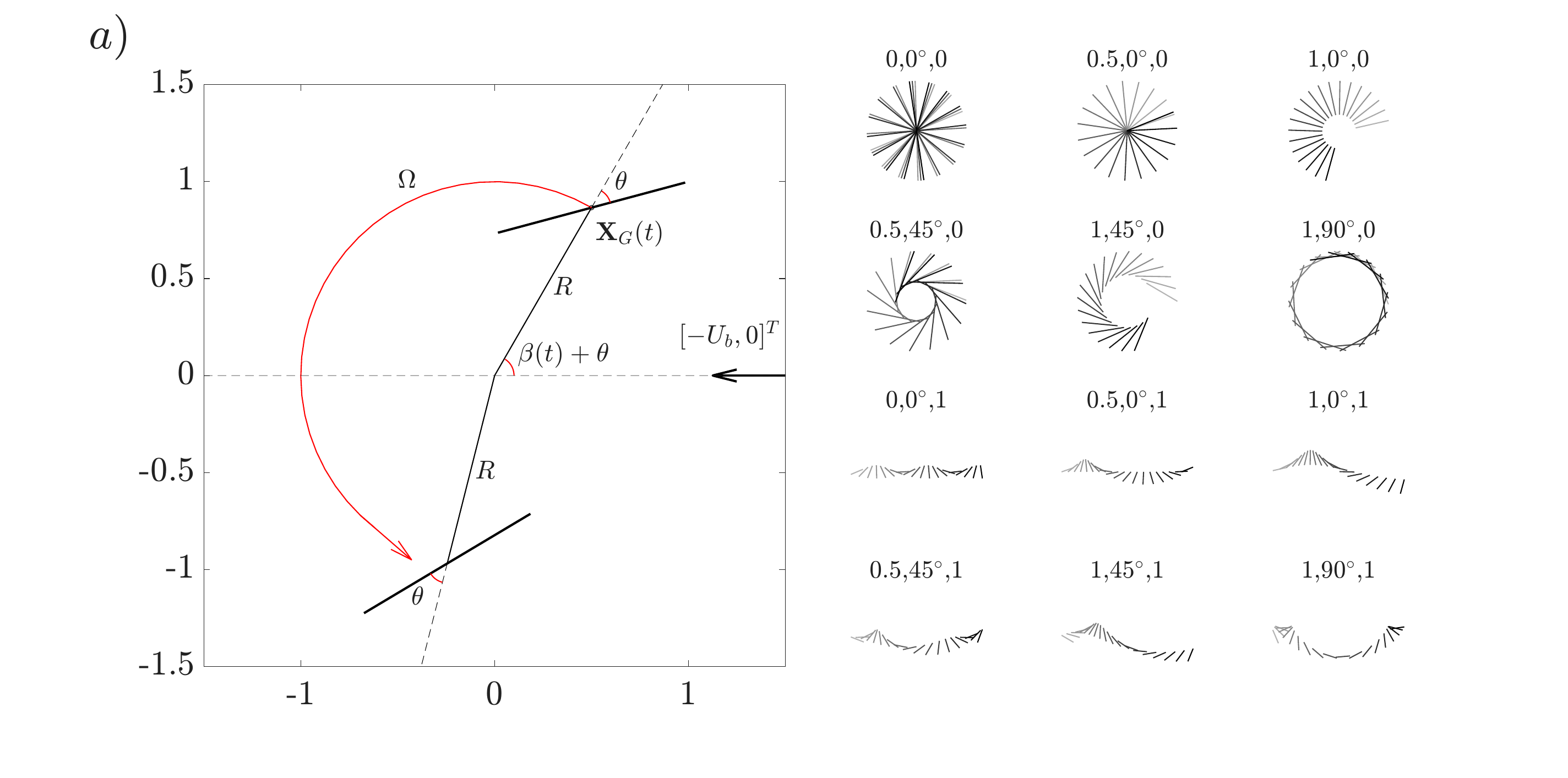}
    \caption{a) Schematic showing the uniformly rotating plate with angular velocity $\Omega$ making angle $\theta$ with the line segment through the origin and its center of mass with length $R$ and background flow $[-U_b,0]^T$. The remaining panels are labeled with a tuple $(R,\theta,U_b)$ and show snapshots of a uniformly rotating plate with the given parameters up to $t = 10$.
    }
    \label{fig:rotating diagram figure}
\end{figure}

We now consider motions with even greater rotation than in the pitch-up example: plates rotating uniformly about the origin. 
More precisely, we consider plates that travel along a circular path
of radius $R$ about the origin and rotate as they do so, such that the angle $\theta$ between the plate and the vector of displacement from the origin remains fixed (see figure \ref{fig:rotating diagram figure}a). For each $(R,\theta)$, we consider the resulting forces and vortex dynamics with and without a background flow $[-U_b,0]^T$. If the background flow is sufficiently strong, the leading and trailing edges swap periodically. The remaining $4\times3$ array of panels in figure \ref{fig:rotating diagram figure} shows the $12$ motions we consider in this section. Each is labeled with a tuple $(R,\theta,U_b)$ and shows snapshots of a plate uniformly rotating with the aforementioned parameters up to $t =10$, the final time we consider in this section. In the panels corresponding to $U_b = 1$ (i.e.~bottom two rows) we show the plate translating rightward with velocity $U_b =1$ which is equivalent to a background flow $[-U_b,0]^T$.

This class of motions is motivated in part by their relevance to vertical axis wind turbines (VAWTs), for which two-dimensional viscous models have been shown to capture aerodynamic forces with reasonable accuracy \cite{lam2016study,nobile2014unsteady}. This is particularly important in configurations involving multiple turbines, where fully resolved simulations can become computationally prohibitive \cite{nazari2018comparison}. In the same spirit, there has been growing interest in low-order models that aim to represent the dominant aerodynamic forces at even lower computational cost, which may enable the computational study of large turbine arrays \cite{nikiforov2022numerical, yuan2020fast}.
\newline

The angular velocity of plate rotation $\Omega$ is chosen as $\Omega = \frac{1}{R^2 + \frac{1}{4} +R\cos(\theta)}$, so that the maximum velocity at the plate tips remains $\approx 1$ as $R$ varies.
More precisely, we set 
\begin{align}
    \boldsymbol{X}_G(t) &= [R\cos(\Omega t), R\sin(\Omega t)]^T + t[U_b,0]^T \\
    \beta(t) &= \Omega t - \theta.
\end{align}
For $\theta = 0^\circ$ we choose $R$ = 0, 0.5, and 1; for $\theta =45^\circ$ we choose $R$ = 0.5 and 1; and for $\theta = 90^\circ$ we choose $R =1$, giving six $(R,\theta)$ pairs in all. Note that the $t[U_b,0]^T$ term corresponds to a background flow speed $U_b$, which we set to 0 and 1 for each $(R,\theta)$ pair, giving $12$ cases in all.
\newline

\begin{figure}[H]
    \centering
    \includegraphics[width=0.85\linewidth, trim = 0 3cm 4cm 0, clip]{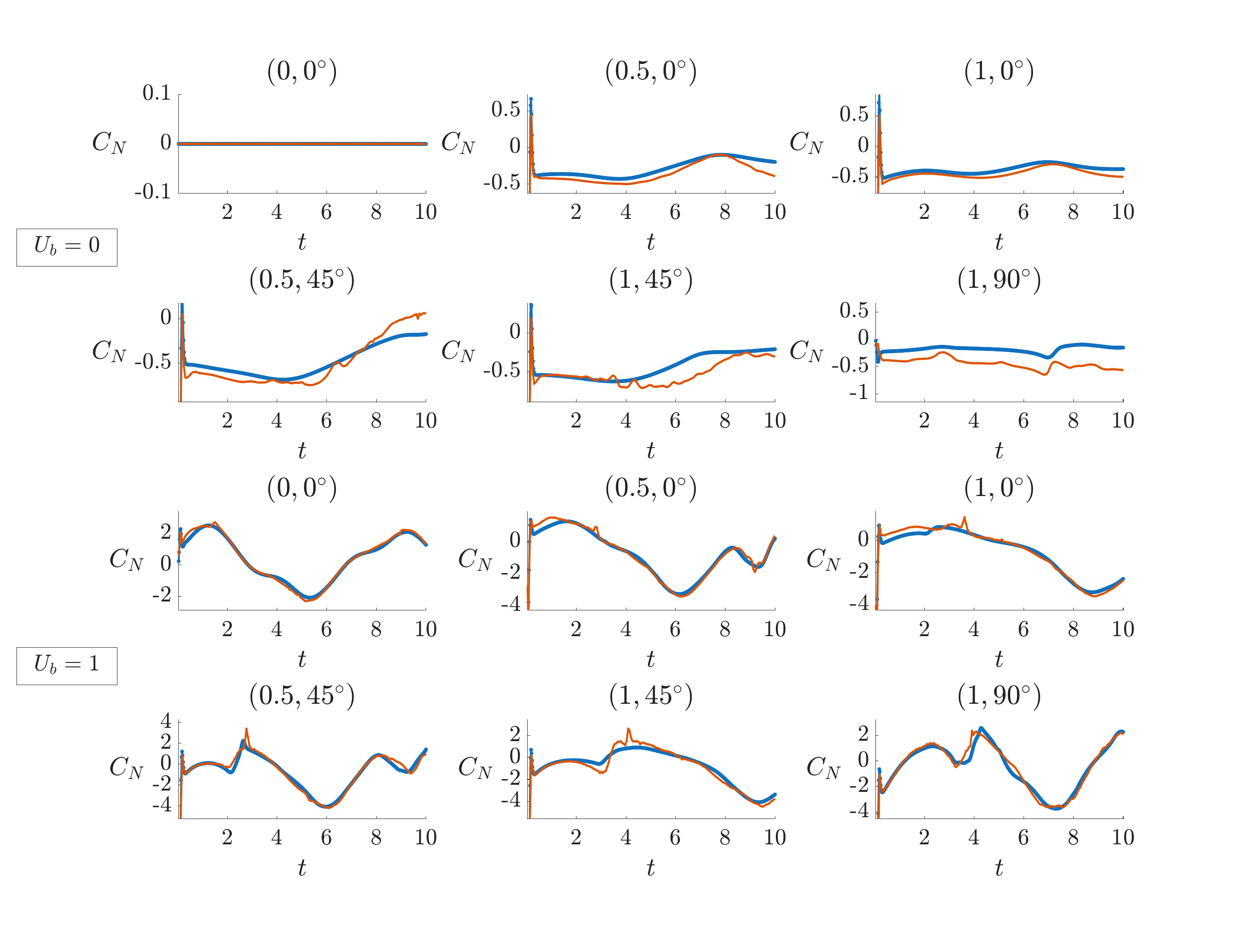}
    \caption{Net normal force of a uniformly rotating plate up to time $t = 10$ for both viscous (blue) and inviscid (orange) models. Each panel is labeled with the corresponding value of  $(R,\theta)$, and the top two and bottom two rows corresponds to the motion with and without a background flow ($-U_b$). }
    \label{spinning force comparison figure}
\end{figure}

\begin{figure}[H]
    \centering
\includegraphics[width=1\linewidth, trim = 3cm 2cm 0cm 1cm, clip]{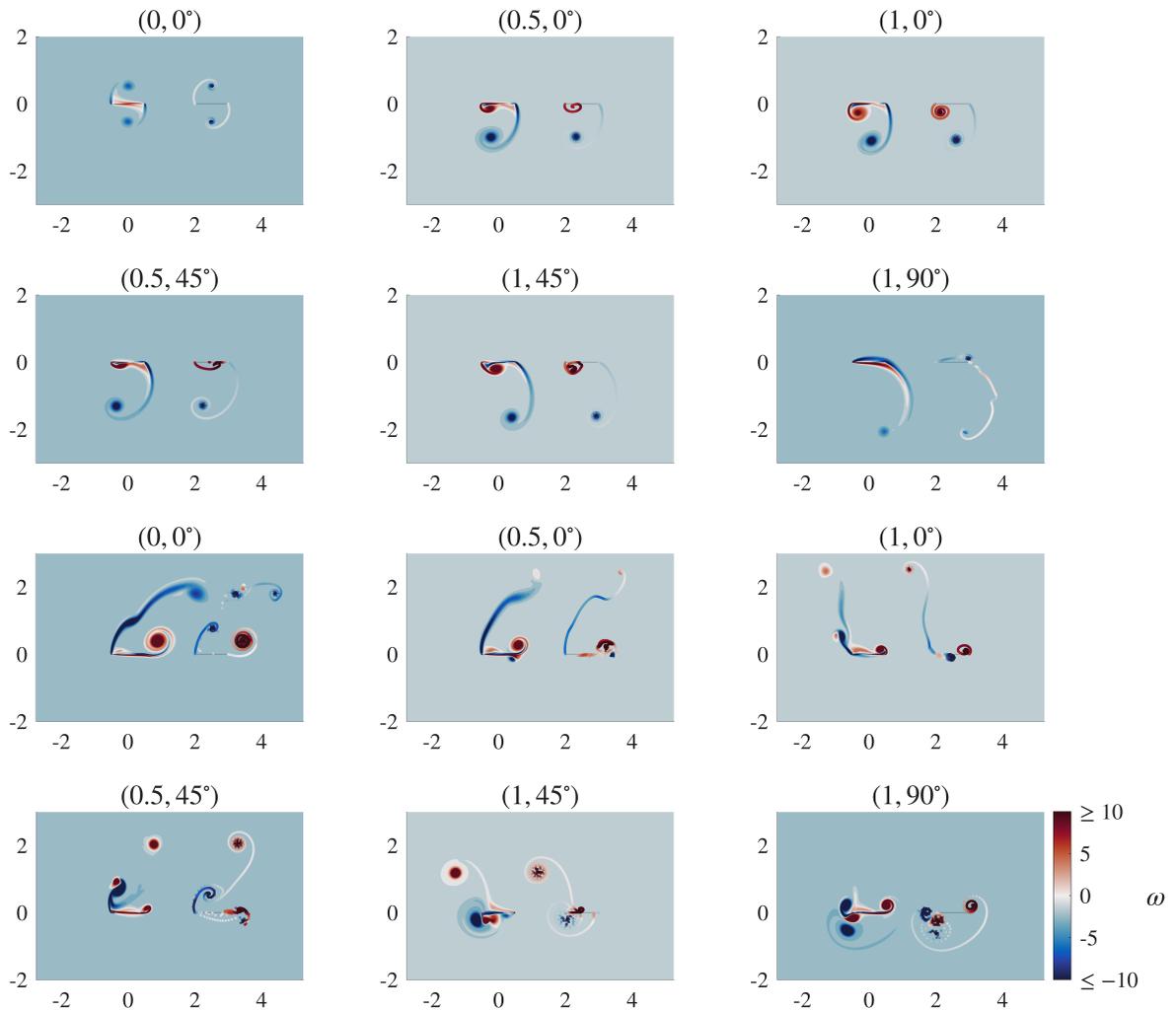}
    \caption{Vorticity $\omega$ at time $t =4$ for a plate rotating with uniform angular velocity about a circle with radius $R$ and making angle $\theta$ with the line connecting the rotation axis to the plate center in inviscid (right) and viscous (left) models. The top two and bottom two rows show the vorticity fields with background flow speeds $U_b = 0$ and 1 respectively. }
    \label{spinning vorticity comparison figure}
\end{figure}

Figure \ref{spinning force comparison figure} presents a $4\times3$ array of plots of net normal force $C_N$, with each panel labeled with the corresponding value of $(R,\theta)$ . The top and bottom two rows correspond to $U_b = 0$ and 1, representing the absence and presence of a background flow respectively. Predictions from the viscous and inviscid models are shown in blue and orange, respectively. A corresponding comparison of the induced vorticity fields at time $t = 4$ is shown in figure \ref{spinning vorticity comparison figure}, with the viscous and inviscid flows shown on the left and right sides of each panel, respectively.
\newline

In a relative sense, the viscous and inviscid forces agree better when $U_b=1$ (the bottom two rows) than when $U_b=0$. In the absence of a background flow, the force magnitudes are approximately an order of magnitude smaller (figure~\ref{spinning force comparison figure}), so in an absolute sense, the level of agreement is similar at $U_b=0$ and 1. Figure~\ref{spinning vorticity comparison figure} shows that the vorticity fields agree slightly better at $U_b=0$ than at $U_b=1$. When a background flow is present, the body motions are more complex, as shown in figure~\ref{fig:rotating diagram figure}, which may lead to more complex vortex dynamics that are more difficult to approximate with the inviscid model. Nonetheless, these results show a generally high level of agreement between the two models.
\newline

The force discrepancies can be traced back to differences in the local vorticity fields near the plate. For example, in figure \ref{spinning force comparison figure} the panel corresponding to $(R,\theta) = (1,45^\circ)$ with $U_b$ = 1 (bottom row, middle column) shows the worst agreement in forces among the examples considered with $U_b = 1$ particularly around $t \approx 4$. Inspecting the corresponding panel in figure \ref{spinning vorticity comparison figure} reveals that at $t = 4$ the left edge of the plate is shedding positive vorticity above the plate rather than below it.
\newline

\begin{figure}[H]
    \centering
\includegraphics[width=1\linewidth, trim = 2cm 0 0cm 0, clip]{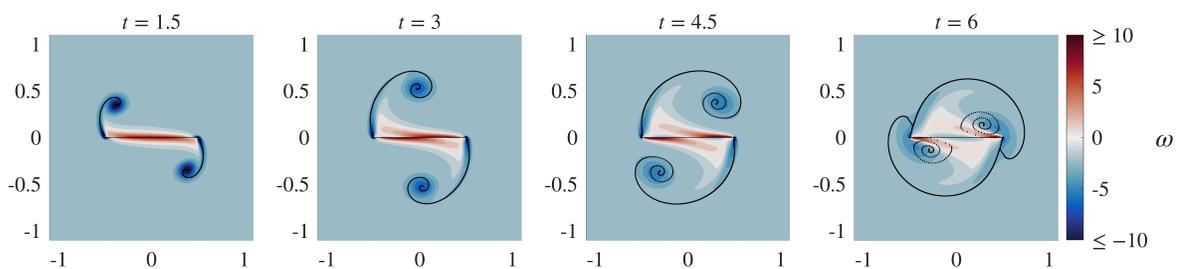}
    \caption{Vorticity $\omega$ at times $t$ = 1.5, 3, 4.5, and 6 for a uniformly rotating plate with $(R,\theta,U_b) = (0,0^\circ,0)$ in a viscous model with the inviscid vortex sheets (black lines) superimposed. }
    \label{spinning example comparison figure}
\end{figure}

When $\theta = 0^\circ$, the rotation axis lies on the line containing the body of the plate. Hence the velocity at any point on the plate body is always perpendicular to its motion, in which case we expect normal (pressure) forces to be strong relative to viscous forces. Thus we may expect good agreement between inviscid and viscous models in the resulting vortex structures. In figure \ref{spinning example comparison figure} we show one such example. The four plots show the vorticity fields of the viscous model at times $t = 1.5,3,4.5,6$ with the outlines of the corresponding inviscid vortex sheets superimposed on them. Even at $t = 6$, when the shed vortices collide with the body of the plate and interact with the viscous boundary layer, the positions of the shed vortex cores show good, though reduced, agreement.
\newline

The robustness and accuracy of the inviscid model in rotary motions with a background flow suggest it may be an effective tool in predicting both the efficiency and vortex shedding patterns of vertical axis wind turbines (VAWT). In \cite{yuan2020fast} for example, a 2D inviscid model similar to that in the present study is coupled with viscous effects to predict the flow induced behind a VAWT.

For the gravity-driven freely-falling plate studied in \cite{loo2025falling}, we observed that beyond a critical density ratio the plate enters an autorotating regime in which the angular velocity grows approximately cubically in time before saturating at a terminal value. While this behavior was identified empirically, the present agreement between inviscid and viscous leading-edge shedding models for rotating plates provides additional evidence for its physical origin. This scaling law can be explained intuitively as follows. Consider the angular momentum balance $I \ddot{\beta} = \tau$
where $I$ is the plate moment of inertia and $\tau$ is the aerodynamic torque. As the plate falls, prior to reaching terminal velocity, the translational velocity satisfies $U(t) \sim g t$. For a thin plate at finite angle of attack, the dominant contribution to the torque arises from pressure forces associated with leading-edge vortex shedding (which scales quadratically with velocity in a quasi-steady approximation), yielding $\tau \sim U^2$. Combining these scalings gives $\ddot{\beta} \sim t^2$, which upon integration implies $\dot{\beta} \sim t^3$, consistent with the cubic growth observed numerically. 

\subsection{Steady translation and the effect of $\delta$}
\label{steady translation subsection}
The preceding examples were all driven primarily by the acceleration of the body. As noted in the previous section, it would appear that in such body-dominated regimes the inviscid model can indeed accurately model both the net normal force and the local vorticity field. We now turn to a flow-dominated case: a plate translating at unit speed in the direction making angle $\theta$ with the $x$-axis. That is, 

\begin{align}
    \boldsymbol{X}_G(t) &= [t\cos(\theta),t\sin(\theta)]^T \\
    \beta(t) &= 0
\end{align}

with $\theta = 10^\circ, \ldots,90^\circ$. After the impulsive start, the plate acceleration vanishes. At the moderate Reynolds number considered ($\mathrm{Re} = 1000$), vortex layer roll-up is a prominent feature of the flow and the resulting forces are dominated by this fluid instability. For bluff bodies, including cylinders and airfoils, such flows are known to exhibit strong sensitivity to the Reynolds number and the period, phase, and amplitude of the forces can change sharply with $\mathrm{Re}$ \cite{yarusevych2009vortex, singh2005vortex, yarusevych2011vortex}. Consequently, in contrast to the previously considered body-dominated examples, one should not expect the same level of agreement between viscous and inviscid models. Unlike the previous examples considered, here we also vary the smoothing parameter $\delta$ which determines the core size of the vortex blobs, as discussed in section \ref{inviscid numerical model subsection}.
\newline

\begin{figure}[H]
    \centering
    \includegraphics[width=0.95\linewidth, trim=2.8cm 4cm 0cm 2cm, clip]{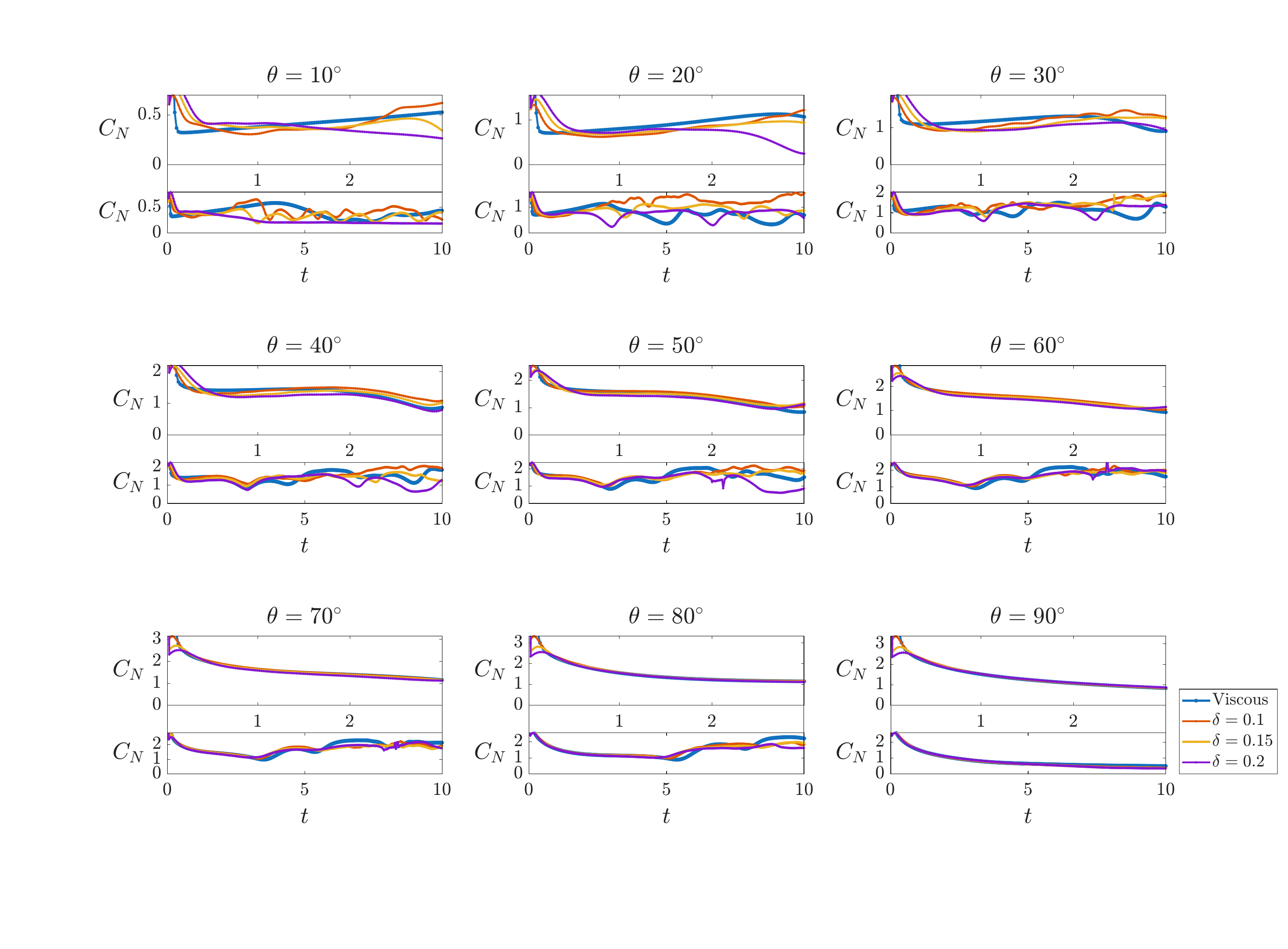}
    \caption{Net normal force induced by an impulsively started plate moving at unit velocity in the direction with angle $\theta = 10^\circ,\ \ldots,90^\circ$  with respect to the $x$ axis up to times $t =3$ (top) and $t = 10$ (bottom).}
    \label{translation force comparison figure}
\end{figure}

\begin{figure}[H]
    \centering
    \includegraphics[width=0.8\linewidth, trim = 1cm 1.5cm 1cm 0.5cm, clip]{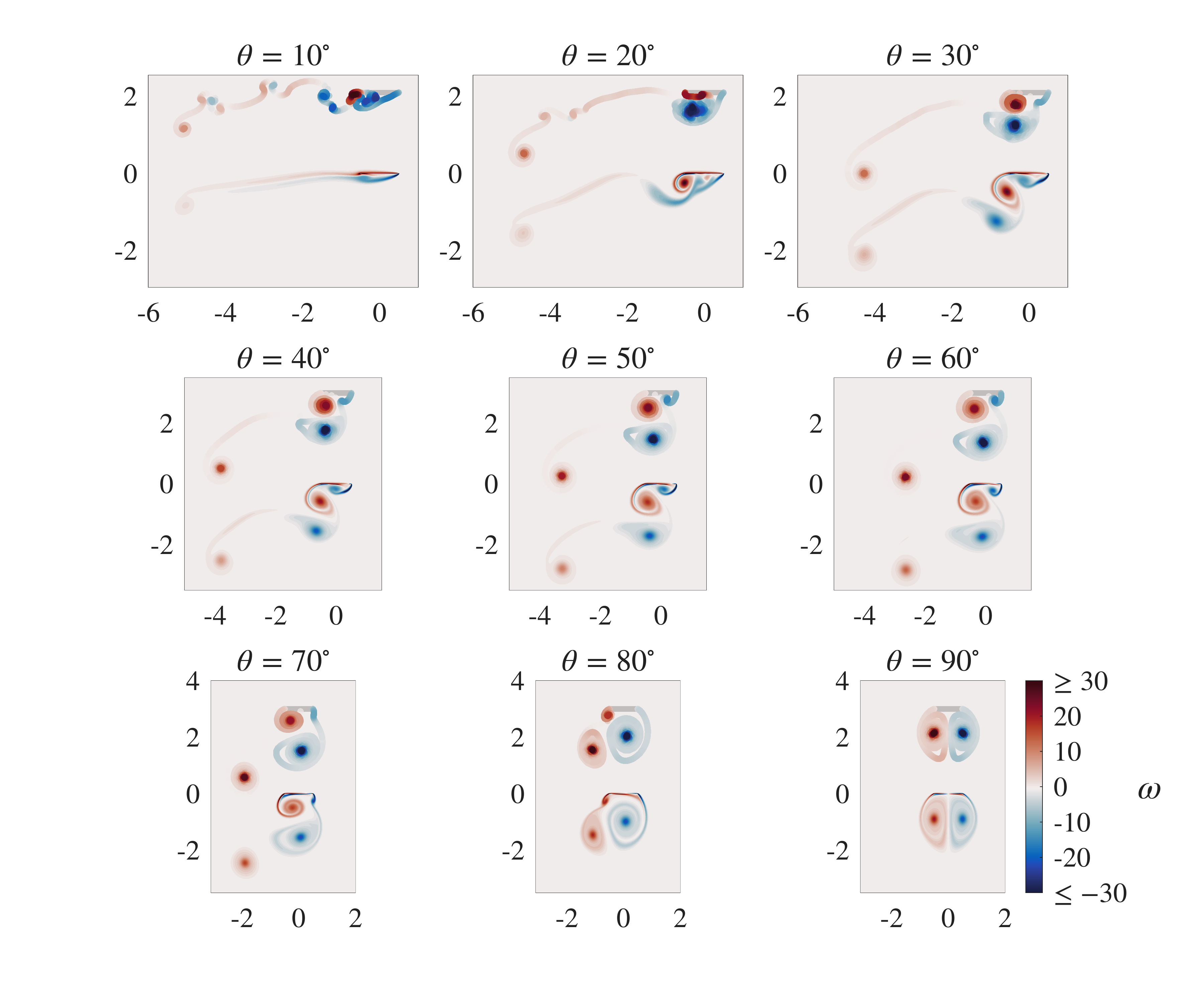}
    \caption{Vorticity $\omega$ at time $t$ = 5 for an impulsively started plate moving at unit velocity in the direction with angle $\theta = 10^\circ,\ \ldots,90^\circ$ with respect to the $x$ axis in inviscid (bottom) and viscous (top) models.}
    \label{viscous translation vorticity comparison figure}
\end{figure}

Figure \ref{translation force comparison figure} shows a $3\times3$ grid of panels showing the net normal force $C_N$ on the plate. The top and bottom subpanels show the net normal force up to times $t =3$ and $t= 10$ respectively. The normal forces on the plates are plotted in blue and orange respectively for the viscous and inviscid ($\delta = 0.1$) models respectively. In addition, we include the normal force induced by the same inviscid model with different values of $\delta$: 0.15 (yellow) and 0.2 (purple) respectively. Figure \ref{viscous translation vorticity comparison figure} shows the corresponding vorticity plots, with the plates placed at $[-1/2,1/2]\times\{0\}$ and at $[-1/2,1/2]\times\{3\}$ for the viscous and inviscid models respectively.  
\newline

One might expect that decreasing the angle of translation increases the discrepancy in the comparison. Indeed, when the plate is aligned with the flow (i.e.~$\theta \approx 0^\circ$), viscous effects dominate. Conversely, when the plate is perpendicular to its direction of motion, pressure effects are approximately maximized relative to viscous effects, leading to closer agreement. Figures \ref{translation force comparison figure} and \ref{viscous translation vorticity comparison figure} appear to confirm this expectation: agreement in both the net normal force and the vorticity plots worsens as $\theta$ decreases. In spite of this, the forces generally agree well for early times as evidenced from the top plot in each panel of figure \ref{viscous translation vorticity comparison figure}. However, the forces eventually deviate due to differences in the vortex shedding phase and structure that accumulate over time. For example, when $10^\circ \leq\theta\leq 30^\circ$ quasi-periodic vortex shedding occurs much earlier in the inviscid model. Viscosity appears to stabilize the vorticity shed from the leading edge and allows it to reattach, delaying the onset of the formation of the leading-edge vortex as shown in the first panel ($\theta = 10^\circ$) of figure \ref{viscous translation vorticity comparison figure}.  In the corresponding panel of figure \ref{translation force comparison figure} the early onset of shedding in the inviscid model introduces high-frequency components to the net normal force not seen in the viscous model. When $\theta = 20^\circ$, however, the opposite occurs.  In the inviscid model, the leading-edge vortex forms a circulation bubble which remains attached for most of the inviscid simulation duration (but detaches around $t= 3$ on the viscous model). This difference in the vortex shedding details leads to comparatively worse agreement here than at other $\theta$ values. The attached leading-edge circulation bubble is visible in the $\theta = 20^\circ$ panel for the top plate (inviscid) of figure \ref{viscous translation vorticity comparison figure}, but is already detaching from the bottom plate (viscous).
\newline

When $\theta$ is large, $\delta$ has little effect on the inviscid model. Indeed, as discussed in section \ref{inviscid numerical model subsection}, tapering $\tilde{\delta}(s)$ to $0$ near the plate in the no-penetration condition (but keeping it uniform during advection) ensures that the total circulation shed at each instant does not depend strongly on $\delta$. When $\theta$ is small however, the close interaction between the body and free sheets makes the effect more apparent. Indeed, the effect of each blob-regularized point vortex is weakened in a $O(\delta)$ radius. Hence, at low angles of attack, increasing $\delta$ has a stabilizing effect similar to viscosity. For example, in the first panel of figure \ref{translation force comparison figure}  ($\theta =10^\circ$), when $\delta = 0.2$, there are no oscillations in the net normal force, as the shed vorticity is relatively weak and the leading-edge vortex has a weak effect on the bound vorticity and forces. 
At higher angles of attack ($\theta \geq 30^\circ$), the effect of changing $\delta$ is much milder, though the discrepancies increase with time as error accumulates. Interestingly, $\delta$ also has a regularizing effect on the impulsive start, with the force singularity smoothed over a time scale $t \approx \delta$. Conversely, as $\delta$ decreases, the impulsive start is more faithfully represented by the inviscid model.  In all cases examined, the inviscid model (with $\delta = 0.1$) remains in good agreement until vortex shedding initiates at $t \approx 3$, as shown in the top panels of Figure~\ref{translation force comparison figure}. This trend aligns with the observation that the model is most reliable for body-dominated flows. 
\newline

\begin{figure}[H]
    \centering
    \includegraphics[width=1\linewidth, trim=0cm 1cm 0cm 1cm, clip]{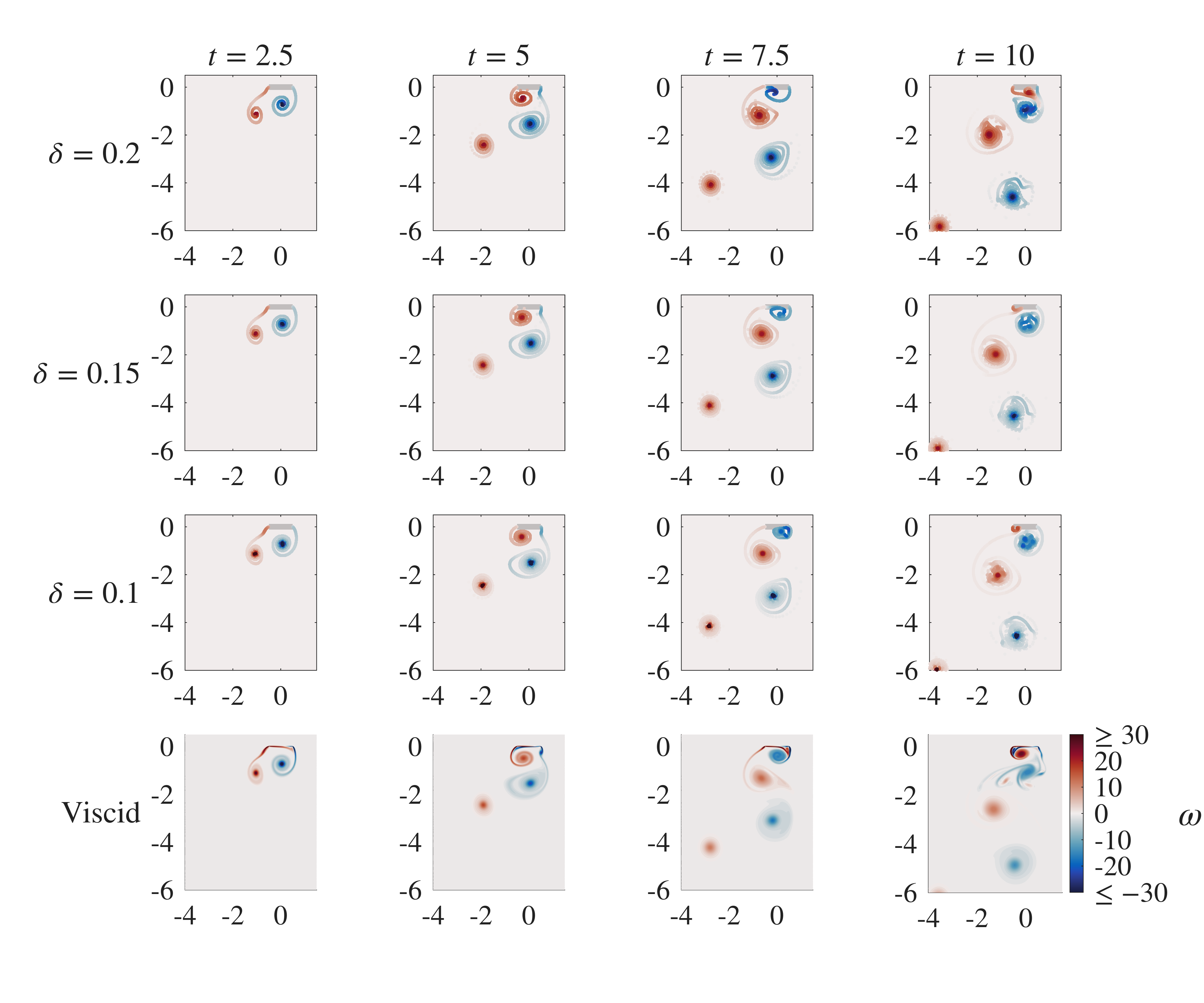}
    \caption{Vorticity $\omega$ at times $t = 2.5,\ 5,\ 7.5,$ and 10 for an impulsively started plate moving at unit velocity at angle $\theta = 70^\circ$ to the $x$ axis in a viscous model and inviscid models with $\delta = 0.1, 0.15,0.2$.}
    \label{translation example comparison figure}
\end{figure}

Even when the shedding patterns are similar, subtle errors can accumulate over time. Figure \ref{translation example comparison figure} demonstrates this effect. It shows a $4\times 4$ grid of vorticity plots corresponding to $\theta = 70^\circ$. The first three rows show the vorticity in the inviscid model for $t$ = 2.5, 5, 7.5, and 10, and $\delta$ = 0.1, 0.15, and 0.2, with the value $0.1$ used in most of this study. The last row shows the same for the Re = 1000 viscous model. As the figure shows, when $t$ is small all inviscid models agree well with the viscous model (e.g.~$t$ = 2.5 and 5). However, at later times (e.g.~$t$ = 7.5 and 10) subtle differences have appeared. For example, in the viscous model, the secondary positive vorticity on the right edge at $t = 7.5$ has rolled up into several smaller positive vortex cores near the right edge. The vortex core shed from the left edge is also noticeably larger than in any of the inviscid models. These discrepancies are smaller in the motions considered in the previous sections, for which the vortex shedding is triggered by body motions rather than natural vortex sheet dynamics, as occurs in quasi-periodic vortex shedding from a plate. Here the dynamics are more sensitive to the change from the viscous to inviscid model.

\section{Summary statistics}
\label{summary statistics section}
In table \ref{tab:L1_error_summary} we list the $L^1$ errors averaged across all the plate motions considered. We define the $L^1$ norm of a time-dependent function $f$ with domain $[0,T]$ as $\|f\|_{L^1} :=\int_0^T|f(t)|dt$ and compute it using the trapezoidal rule. In the previous sections, we compared $C_N(t)$, the net normal force on the plate as a function of time, between the viscous and inviscid models. We define the relative $L^1$ error between the inviscid and viscous normal forces as 
\begin{equation}
    \frac{2\|C_N^{viscous} - C_N^{inviscid}\|_{L^1}}{\|C_N^{viscous} \|_{L^1} + \|C_N^{inviscid}\|_{L^1}}
\end{equation}

This quantity provides a measure of the accuracy of the inviscid models in predicting the Re = 1000 results over different types of motions. For the motions we defined as flow-dominated---steady translation and uniform rotation without background flow---the maximum $L^1$ error exceeds $50\%$. For the body-dominated motions, the average $L^1$ errors are between $11-19\%$.
\begin{table}[H]
\centering
\caption{Relative $L^1$ error in the net normal force between inviscid vortex sheet simulations and Navier--Stokes simulations at Re = 1000 for each class of plate motion. The $L^1$ norm is computed over the full simulation time interval.}
\vspace{0.5cm}
\begin{tabular}{lccc}
\hline
\textbf{Motion type} 
& \textbf{Min. $L^1$ error} 
& \textbf{Avg. $L^1$ error} 
& \textbf{Max. $L^1$ error} \\
\hline
Oscillating plates                              & 0.092 & 0.18 & 0.29 \\
Heaving plates                                  & 0.063 & 0.19 & 0.30 \\
Pitching plates                                 & 0.097 & 0.18 & 0.25 \\
Mixed heaving and pitching                      & 0.13 & 0.18 & 0.21 \\
Pitch-up motions                                & 0.084 & 0.11 & 0.17 \\
Uniform rotation with background flow           & 0.083 & 0.15 & 0.21 \\
Uniform rotation without background flow        & 0.15 & 0.31 & 0.84 \\
Steady translation                              & 0.10 & 0.20 & 0.51 \\
\hline
\end{tabular}
\label{tab:L1_error_summary}
\end{table}

\section{Conclusion}
\label{conclusion section}
In this work, we performed a systematic comparison between an inviscid vortex sheet model with continuous leading-edge shedding and direct Navier--Stokes simulations for a broad range of unsteady plate motions at moderate Reynolds number ($\mathrm{Re} = 1000$). We assessed the extent to which inviscid formulations can reproduce viscous force predictions and the associated vortex dynamics. Our results show that in body-dominated regimes, where the forces are primarily driven by prescribed plate accelerations, the inviscid model captures both the normal force histories and the qualitative structure of the induced vorticity field near the plate with good accuracy. In these cases, the vortex shedding patterns near the edges of the plate are represented well by the vortex sheet formulation, and discrepancies attributable to viscous effects remain comparatively small. By contrast, for flow-dominated dynamics (i.e.~with steady body motions), agreement between viscous and inviscid force predictions is reduced at low angles of attack. At $\mathrm{Re} = 1000$, such configurations exhibit vortex shedding that is sensitive to the detailed near-body flow, a sensitivity that is well-documented for bluff bodies. In these regimes, small differences in vortex placement and evolution can lead to appreciable differences in the resulting forces. While the inviscid model continues to capture overall force trends, quantitative agreement is necessarily limited by the absence of viscous dissipation and Reynolds-number-dependent effects. 
\newline

The central outcome of this study is that allowing for continuous leading-edge vortex shedding enables stable, robust, and generally accurate inviscid simulations across a wide range of unsteady motions. These results clarify the regimes in which inviscid vortex sheet models can be expected to provide accurate force predictions, as well as the regimes in which their limitations are more apparent. The findings support the use of such models as efficient tools for interpreting force generation mechanisms and exploring large parameter spaces in unsteady high-Reynolds-number flows, while also underscoring the importance of viscous simulations for capturing flow-dominated dynamics.

\section*{Acknowledgments}
This research was supported by the NSF-DMS Applied Mathematics program under
award number DMS-2204900.

\section*{Declaration of interests}
The authors report no conflict of interest.
\appendix

\section{Navier--Stokes equations in a translating and rotating reference frame}
\label{navier stokes in body frame derivation appendix}

In this section we give a detailed derivation of the Navier--Stokes equations in a translating and rotating frame of reference. The main ideas can be found in \cite{li2002moving} and in standard textbooks on classical mechanics, such as \cite{morin2008introduction}, but we include the derivation here for completeness and to fix notation.

\subsection{Pure translation}

We first consider the simpler case of a purely translating, non-rotating frame of reference. Let $\boldsymbol{R}(t)$ denote the position of the origin of this translating frame with respect to an absolute stationary frame, and define
\[
\boldsymbol{U}(t) := \dot{\boldsymbol{R}}(t), 
\qquad 
\boldsymbol{A}(t) := \dot{\boldsymbol{U}}(t).
\]

Let $\boldsymbol{r}$ denote the position of a fluid parcel in the absolute frame, and let $\boldsymbol{r}'$ denote its position in the translating frame. Then
\[
\boldsymbol{r}' := \boldsymbol{r} - \boldsymbol{R},
\qquad
\boldsymbol{r} = \boldsymbol{r}' + \boldsymbol{R}.
\]

Let $\boldsymbol{u}$ denote the velocity of the fluid parcel in the absolute frame and let $\boldsymbol{v}$ denote its velocity in the translating frame. Differentiating $\boldsymbol{r} = \boldsymbol{r}' + \boldsymbol{R}$ with respect to time yields
\begin{equation}
\boldsymbol{u}
= \frac{d}{dt}\boldsymbol{r}
= \frac{d}{dt}\boldsymbol{r}' + \dot{\boldsymbol{R}}
= \boldsymbol{v} + \boldsymbol{U}.
\label{translation velocity relation}
\end{equation}

Differentiating once more gives
\begin{equation}
\frac{d}{dt}\boldsymbol{u}
= \frac{d}{dt}\boldsymbol{v} + \boldsymbol{A}.
\label{translation acceleration relation}
\end{equation}

The Navier--Stokes equations in the absolute stationary frame are
\begin{equation}
\frac{d}{dt}\boldsymbol{u}
=
-\nabla p + \frac{1}{\mathrm{Re}}\Delta \boldsymbol{u}.
\label{navier stokes lab}
\end{equation}
Since $\boldsymbol{U}(t)$ is spatially uniform, we have $\Delta \boldsymbol{u} = \Delta \boldsymbol{v}$. Substituting \eqref{translation acceleration relation} into \eqref{navier stokes lab} and relabeling $\boldsymbol{v}$ as $\boldsymbol{u}$ and $\boldsymbol{r}'$ as $\boldsymbol{r}$ yields
\begin{equation}
\frac{d}{dt}\boldsymbol{u}
=
-\nabla p + \frac{1}{\mathrm{Re}}\Delta \boldsymbol{u}
-\boldsymbol{A}.
\label{navier stokes translating}
\end{equation}
This is the Navier--Stokes equation in a purely translating, non-rotating frame.

\subsection{Mixed translation and rotation}

We now consider a rotating frame attached to the body. Let $\boldsymbol r(\boldsymbol{x},t)$ denote the position of a fluid parcel originating from position $\boldsymbol{x}$ relative to the translating frame discussed above. Let $\mathcal B = \{\boldsymbol{b}_1,\boldsymbol{b}_2,\boldsymbol{b}_3\}$ denote an orthonormal frame rotating with angular velocity $\boldsymbol{\Omega}(t)$ relative to the translating (non-rotating) frame introduced above. As before, the translating frame has velocity $\boldsymbol{U}(t)$ and acceleration $\boldsymbol{A}(t) = \dot{\boldsymbol{U}}(t)$ relative to an absolute stationary frame.
\newline

Notice that observers in the frame $\mathcal B$ regard the basis vectors $\boldsymbol{b}_i$ as fixed in time, while observers in the translating frame see them as time-dependent. Consequently, observers in the two frames must define different time derivatives. Let $\frac{d}{dt}^{\mathcal B}$ denote the time derivative taken in the rotating frame $\mathcal B$, defined by the requirement that
\[
\frac{d}{dt}^{\mathcal B}\boldsymbol{b}_i = \boldsymbol{0},
\qquad i=1,2,3,
\]
(ie. The frame $\mathcal{B}$ is constant with respect to this derivative). Explicitly, for any vector
\[
\boldsymbol{B} = B^1\boldsymbol{b}_1 + B^2\boldsymbol{b}_2 + B^3\boldsymbol{b}_3,
\]
we define
\[
\frac{d}{dt}^{\mathcal B}\boldsymbol{B}
:= \dot{B}^1\boldsymbol{b}_1 + \dot{B}^2\boldsymbol{b}_2 + \dot{B}^3\boldsymbol{b}_3.
\]
The derivative $\frac{d}{dt}$ without a superscript is defined analogously, and so denotes the standard time derivative in the translating frame.
\newline

Suppose for the moment that the following identity holds:
\begin{equation}
\frac{d}{dt}^{\mathcal B}
=
\frac{d}{dt}
-
\boldsymbol{\Omega}\times(\cdot).
\label{derivative in new frame equation}
\end{equation}

Let $\boldsymbol{u}$ denote the velocity of a fluid parcel in the translating frame, and let $\boldsymbol{v}$ denote its velocity as measured in the rotating frame $\mathcal B$, defined by
\[
\boldsymbol v := \frac{d}{dt}^{\mathcal B}\boldsymbol r,
\qquad
\boldsymbol u := \frac{d}{dt}\boldsymbol r.
\]
Then \eqref{derivative in new frame equation} implies
\begin{equation}
\boldsymbol{u}
=
\boldsymbol{v}
+
\boldsymbol{\Omega}\times\boldsymbol{r},
\label{u minus v equation}
\end{equation}
and hence
\[
\boldsymbol{u}-\boldsymbol{v}
=
\boldsymbol{\Omega}\times\boldsymbol{r}.
\]

Differentiating once more gives
\begin{align}
\frac{d}{dt}\boldsymbol{u}
&=
\Big(\frac{d}{dt}^{\mathcal B} + \boldsymbol{\Omega}\times\cdot\Big)
\Big(\boldsymbol{v} + \boldsymbol{\Omega}\times\boldsymbol{r}\Big) \\
&=
\frac{d}{dt}^{\mathcal B}\boldsymbol{v}
+2\,\boldsymbol{\Omega}\times\boldsymbol{v}
+\dot{\boldsymbol{\Omega}}^{\mathcal B}\times\boldsymbol{r}
+\boldsymbol{\Omega}\times(\boldsymbol{\Omega}\times\boldsymbol{r}),
\end{align}
where $\dot{\boldsymbol{\Omega}}^{\mathcal B}$ denotes the time derivative of $\boldsymbol{\Omega}$ taken in the rotating frame. Thus
\begin{equation}
\frac{d}{dt}\boldsymbol{u}
=
\frac{d}{dt}^{\mathcal B}\boldsymbol{v}
+2\,\boldsymbol{\Omega}\times\boldsymbol{v}
+\dot{\boldsymbol{\Omega}}^{\mathcal B}\times\boldsymbol{r}
+\boldsymbol{\Omega}\times(\boldsymbol{\Omega}\times\boldsymbol{r}).
\label{dtu equation}
\end{equation}

By the chain rule,
\[
\frac{d}{dt}^{\mathcal B}\boldsymbol{v}
=
\frac{\partial}{\partial t}^{\mathcal B}\boldsymbol{v}
+
(\boldsymbol{v}\cdot\nabla)\boldsymbol{v}.
\]
The gradient operator is frame-independent, and the Laplacian is invariant under orthonormal changes of coordinates.

Since the translating frame satisfies \eqref{navier stokes translating}, the Navier--Stokes equations take the form
\begin{equation}
\frac{d}{dt}\boldsymbol{u}
=
-\nabla p
+
\frac{1}{\mathrm{Re}}\Delta\boldsymbol{u}
-
\boldsymbol{A}.
\label{navier stokes appendix}
\end{equation}
Using \eqref{u minus v equation} and the fact that $\Delta(\boldsymbol{\Omega}\times\boldsymbol{r})=0$, we may replace $\Delta\boldsymbol{u}$ by $\Delta\boldsymbol{v}$.

Substituting \eqref{dtu equation} into \eqref{navier stokes appendix} then yields
\begin{align}
\frac{\partial}{\partial t}^{\mathcal B}\boldsymbol{v}
+(\boldsymbol{v}\cdot\nabla)\boldsymbol{v}
+2\,\boldsymbol{\Omega}\times\boldsymbol{v}
+\dot{\boldsymbol{\Omega}}^{\mathcal B}\times\boldsymbol{r}
+\boldsymbol{\Omega}\times(\boldsymbol{\Omega}\times\boldsymbol{r})
=
-\nabla p
+
\frac{1}{\mathrm{Re}}\Delta\boldsymbol{v}
-
\boldsymbol{A}.
\end{align}

Finally, treating $\mathcal B$ as the canonical basis (suppressing the superscript $\mathcal B$ and relabeling $\boldsymbol{v}$ as $\boldsymbol{u}$), we obtain
\begin{equation}
\partial_t\boldsymbol{u}
+(\boldsymbol{u}\cdot\nabla)\boldsymbol{u}
-\frac{1}{\mathrm{Re}}\Delta\boldsymbol{u}
=
-\nabla p
-2\,\boldsymbol{\Omega}\times\boldsymbol{u}
-\dot{\boldsymbol{\Omega}}\times\boldsymbol{r}
-\boldsymbol{\Omega}\times(\boldsymbol{\Omega}\times\boldsymbol{r})
-\boldsymbol{A}.
\end{equation}
This is exactly equation \eqref{unsimplified Navier Stokes equation in the body frame of the plate}. The additional Coriolis, Euler, and centrifugal acceleration terms all originate from the time dependence of the rotating basis through the identity \eqref{derivative in new frame equation}. To complete the derivation it remains to show equation \eqref{derivative in new frame equation}. To see this, we claim that for $i = 1, 2, 3$ we have
\begin{equation}
    \frac{d}{dt}\boldsymbol{b}_i = \boldsymbol{\Omega}\times\boldsymbol{b}_i
\end{equation}
Note that this equation immediately follows from the assumption that the frame $\mathcal{B}$ rotates with angular velocity $\boldsymbol{\Omega}$. Indeed, this follows from the following two observations. Firstly, the distance from $\boldsymbol{b}_i$ to the axis of rotation (i.e.~the line through $\boldsymbol{\Omega}$) is $|\boldsymbol{b}_i|\sin(\theta_i) = \sin(\theta_i)$, where $\theta_i$ is the angle $\boldsymbol{b}_i$ makes with $\boldsymbol{\Omega}$. Hence, as the angular velocity of this point is $|\boldsymbol{\Omega}|$ the magnitude of $\frac{d}{dt}\boldsymbol{b}_i$ is thus $|\boldsymbol{\Omega}|\sin(\theta_i)$. Secondly, as the point is rotating about the line containing $\boldsymbol{\Omega}$, it must be moving perpendicular to the plane containing both $\boldsymbol{\Omega}$ and $\boldsymbol{b}_i$. Hence by the right hand rule, $\frac{d}{dt}\boldsymbol{b}_i$ must be a vector proportional to $\widehat{\boldsymbol{\Omega}\times\boldsymbol{\boldsymbol{b}_i}}$ where the hat denotes the vector has been normalized. Combining these two observations yields

\begin{equation}
    \frac{d}{dt}\boldsymbol{b}_i = |\boldsymbol{\Omega}|\sin(\theta_i)\widehat{\boldsymbol{\Omega}\times\boldsymbol{b}_i} = \boldsymbol{\Omega}\times\boldsymbol{b}_i.
\end{equation}

Equation \eqref{derivative in new frame equation} now follows immediately. Indeed, for any 
$\boldsymbol{B} = B^1\boldsymbol{b}_1 + B^2\boldsymbol{b}_2 + B^3\boldsymbol{b}_3$,
\begin{align}
    \frac{d}{dt}\boldsymbol{B}&= \frac{d}{dt}\bigg(B^1\boldsymbol{b}_1 + B^2\boldsymbol{b}_2 + B^3\boldsymbol{b}_3\bigg) \\ 
    &=\dot{B}^1\boldsymbol{b}_1 + \dot{B}^2\boldsymbol{b}_2 + \dot{B}^3\boldsymbol{b}_3 + B^1\dot{\boldsymbol{b}}_1 + B^2\dot{\boldsymbol{b}}_2 + B^3\dot{\boldsymbol{b}}_3 \\
    &=\dot{B}^1\boldsymbol{b}_1 + \dot{B}^2\boldsymbol{b}_2 + \dot{B}^3\boldsymbol{b}_3 + \boldsymbol{\Omega}\times\big(B^1\boldsymbol{b}_1 + B^2\boldsymbol{b}_2 + B^3\boldsymbol{b}_3\big) \\ 
    &= \frac{d}{dt}^\mathcal{B} \boldsymbol{B} + \boldsymbol{\Omega}\times \boldsymbol{B}.
\end{align}

\section{Numerical implementation of the Kutta condition}
\label{kutta condition appendix}
The Kutta condition is the principle that the fluid velocity should be bounded at every point in space and at every instant. As discussed in \cite{loo2025falling} this is mathematically equivalent to requiring the vortex strength $\gamma$ to be a continuous function across the edges of the plate:
\begin{equation}
    \gamma_b(\pm1/2,t) = \gamma_\pm(\pm1/2,t)
\end{equation}

In this appendix we briefly discuss how one can enforce this Kutta condition in step 4 of the numerical algorithm \ref{full numerical algorithm}. We find that enforcing the Kutta condition this way improves the accuracy and stability of the numerical model. The no-penetration condition can be written in the form 

\begin{equation}
    C[\gamma_b](s,t)=\int_{-1/2}^{1/2} \tilde{C}(\boldsymbol{X}(s,t) - \boldsymbol{X}(s',t)\gamma_b(s',t)ds' = f(s,t)
    \label{cauchy equation}
\end{equation}
where $$\tilde{C}(\boldsymbol{x}) = \frac{1}{2\pi}\hat{\boldsymbol{n}}\cdot\boldsymbol{K}(\boldsymbol{x)}$$ and $$f(s,t) = \hat{\boldsymbol{n}}\cdot\boldsymbol{U} - \frac{1}{2\pi}\hat{\boldsymbol{n}}\cdot\bigg(\int_{0}^{\Gamma_-(t)}\boldsymbol{K}_{\tilde{\delta}(s')}\big(\boldsymbol{X}(s,t) - \boldsymbol{X_-}(\Gamma',t)\big)d\Gamma' + \\ \int_{0}^{\Gamma_+(t)}\boldsymbol{K}_{\tilde{\delta}(s')}\big(\boldsymbol{X}(s,t) - \boldsymbol{X}_+(\Gamma',t)\big)d\Gamma'\bigg) $$

Generically, the solution of \eqref{cauchy equation} has inverse-square-root singularities at the edges \cite{loo2025falling}. Hence $\gamma_b(s,t)  = \frac{\sigma(s,t)}{\sqrt{1/4 - s^2}}$. Now one may change variables using $s = -\frac{1}{2}\cos(\alpha)$ so that we have 

\begin{equation}
    \int_{0}^{\pi} \tilde{C}(\boldsymbol{X}(\alpha,t) - \boldsymbol{X}(\alpha',t)\sigma(\alpha',t)d\alpha' = f(\alpha,t)
    \label{cauchy equation sigma}
\end{equation}

We determine $\sigma$ on the plate using the collocation method \cite{Golberg1990IntroductionTT}. Let $\alpha_j = j\frac{\pi}{n}$ with $j = 0,\ \ldots,n$ be a uniform grid over $[0,\pi]$, and $\alpha_{j + 1/2}  = (j + 1/2)\frac{\pi}{n}$ with $j = 0,\ \ldots,n-1 $ be the midpoints of the uniform grid. Let $w_j,\ j = 0,\ \ldots,n$ be quadrature weights on the uniform grid. Let $s_{j} = -\frac{1}{2}\cos(\alpha_j)$.

We may hence discretize equation \eqref{cauchy equation sigma} to form an $n\times (n+1)$ under-determined linear system of equations

\begin{equation}
    \sum_{j=0}^n \tilde{C}(\boldsymbol{X}(\alpha_{k+1/2},t) - \boldsymbol{X}(\alpha_j,t))\sigma(\alpha_j,t)w_j = f(\alpha_{k+1/2},t)
    \label{cauchy equation sigma discretized}
\end{equation}
with $k = 0,\ldots, n-1$. Recall that Kelvin's circulation theorem demands that $$ \int_{-1/2}^{1/2}\gamma_b(s,t)ds  +\Gamma_+(t) + \Gamma_-(t) = 0 .$$ After changing variables again, Kelvin's theorem is discretized as $$\sum_{j =0}^n \sigma(\alpha_j,t)w_j = -\Gamma_+(t) - \Gamma_-(t).$$ 
Now we have $n+1$ equations and $n+1$ unknowns, so we may invert the resulting linear system to obtain $\sigma(\alpha_j,t)$ for $j = 0,\ldots, n$. For $\gamma_b(\pm1/2,t)$ to be finite we necessarily require $\sigma(\pm1/2,t) = 0$. Hence we set $\sigma(\alpha_j,t) = 0$ for $j = 0$ and $n$. Finally, we obtain $\gamma_b$ from $\sigma$ through the observation that $$\int_0^\alpha\sigma(\alpha',t)d\alpha' = \Gamma_b(\alpha,t) = \Gamma_b(s,t)=\int_0^s\gamma_b(s',t)ds'$$
with $s = -\frac{1}{2}\cos(\alpha)$. Hence $$\frac{d}{ds}\int_0^{\alpha} \sigma(\alpha',t) d\alpha' = \gamma_b(s,t).$$
Discretizing the derivative and integral on the left-hand side yields $\gamma_b(s_j,t),\ j = 0,\ \ldots,n$ on the right-hand side, including $\gamma_b(\pm1/2,t)$ for use in the Kutta condition.

\section{Reynolds-number sensitivity}

\label{reynolds number sensitivity appendix}
All comparisons presented in this study were performed at $\mathrm{Re}=1000$, which corresponds to an object of length $\approx$ 3 cm in water, or 12 cm in air, moving at about one body length per second. Generally, good agreement was obtained. To assess whether the observed agreement is specific to this Reynolds number, we perform an additional comparison for a representative class of motions at three Reynolds numbers.
\newline

\begin{figure}[H]
    \centering
    \includegraphics[width=1\linewidth]{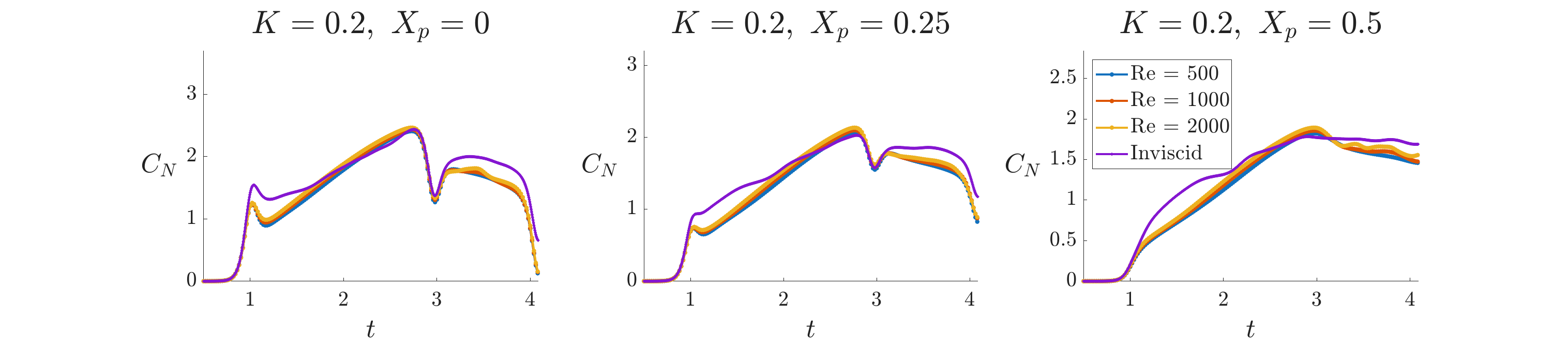}
    \caption{Induced forces of a pitch-up maneuver with $X_p = 0,0.25,0.5$ and $KC = 0.2$ for $\textrm{Re} = 500,1000,2000$.}
    \label{fig:reynolds number sensitivity figure}
\end{figure}

Figure \ref{fig:reynolds number sensitivity figure} compares the normal force history predicted by the inviscid vortex sheet model and direct Navier–Stokes simulations for a pitch-up maneuver with $K = 0.2$ and pivot point $X_p$ = 0, 0.25, and 0.5 at $\mathrm{Re}=500$, $1000$, and $2000$. Across this range, the dimensionless net normal force $C_N$ does not vary much in the viscous model, and is close to that of the inviscid model. Table \ref{tab:re comparison table} shows the $L^1$ error of the inviscid model relative to the Navier-Stokes simulation defined in section \ref{summary statistics section}. The figure and table show that the effect of Re is slight in this range; the error decreases by about 1--2\% with each doubling of $\textrm{Re}$.

\begin{table}[H]
\centering
\caption{Relative $L^1$ error between inviscid and viscous models undergoing a pitch-up maneuver with different pivot positions $X_p$ and Reynolds numbers $Re$.}
\vspace{0.5cm}
\label{tab:re comparison table}
\begin{tabular}{c|ccc}
\hline
$X_p$ & $Re=500$ & $Re=1000$ & $Re=2000$ \\
\hline
0    & 0.11 & 0.098 & 0.091 \\
0.25 & 0.11 & 0.094 & 0.086 \\
0.5  & 0.12 & 0.096 & 0.079 \\
\hline
\end{tabular}
\end{table}

These results indicate the possibility of a broad range of Re in which inviscid vortex sheet models with continuous leading-edge shedding may provide useful force predictions.
\printbibliography[
heading=bibintoc,
title={References}
]
\end{document}